# Toward the theoretically observable limit of electron density distribution by single-crystal synchrotron X-ray diffraction: The case of orbitally ordered Ti-3$d^1$ in YTiO$_3$

Authors

**Terutoshi Sakakura[a]\*, Yoshihisa Ishikawa[b], Shunji Kishimoto[c], Yasuyuki Takenaka[d], Kiyoaki Tanaka[ef], Shigeki Miyasaka[g], Yoshinori Tokura[hi], Yukio Noda[ac], Nobuo Ishizawa[j], Hajime Sagayama[c], Hajime Yamamoto[a] and Hiroyuki Kimura[a]**

[a] Institute of Multidisciplinary Research for Advanced Materials, Tohoku University, Sendai, Miyagi 980-8577, Japan

[b] Neutron Science and Technology Center, Comprehensive Research Organization for Science and Society, Tokai, Naka, Ibaraki 319-1106, Japan

[c] Institute of Materials Structure Science, High Energy Accelerator Research Organization, Tsukuba, Ibaraki 305-0801, Japan

[d] Hokkaido University of Education at Hakodate, Hakodate, Hokkaido 040-8567, Japan

[e] Research Division, Nagoya Industrial Science Research Institute, Nagoya, Aichi 464-0819, Japan

[f] Graduate School of Engineering, Nagoya Institute of Technology, Gokiso-cho, Showa-ku, Nagoya, Aichi 466-8555, Japan

[g] Department of Physics, Graduate School of Science, Toyonaka, Osaka 560-0043, Japan

[h] Department of Applied Physics, The University of Tokyo, Bunkyo-ku, Tokyo 113-8656, Japan

[i] Center for Emergent Matter Science, RIKEN, Wako, Saitama 351-0198, Japan

[j] Advanced Ceramics Research Center, Nagoya Institute of Technology, Tajimi, Gifu 507-0051, Japan

Correspondence email: terutoshi.sakakura.a7@tohoku.ac.jp

**Funding information**     Japan Society for the Promotion of Science (grant nos. 21244051; 23654098; 24340064; 15H02038); High Energy Accelerator Research Organization (grant no. Quantum Beam Sakakura).

By reviewing the mathematics of X-ray crystal structure analysis, we discuss how to asymptotically approach the theoretical limit of the observable electron density distribution in real experiments.

**Abstract**   The theoretically observable limit of electron density distribution by single-crystal X-ray diffraction is discussed. When $F_{orb}$ and $\delta F$ are defined as, respectively, the partial structure factor for an orbital and the deviation of the observed $F$ from the true $F$, the accuracy of electron density attributable to $F_{orb}$ is chiefly determined by the number of reflections satisfying the condition $F_{orb}/F > \delta F/F$. Since $F_{orb}/F$, which is generally small for crystals with large $F(0,0,0)$, is constant under a given set of experimental conditions, $\delta F/F$ must be reduced to increase the number of reflections satisfying $F_{orb}/F > \delta F/F$. The present paper demonstrates how to reduce $\delta F$ mathematically and experimentally, and the following topics are covered: the Poisson statistics, accumulation of errors in the data collection and reduction procedure, multiple diffraction, conversion error from $F^2$ to $F$ in refinement programs, which is unavoidable when the input quantities have different dimension from $F$, weighting of reflections, and tips. For demonstration, observation of the electron density of the Ti-$3d^1$ orbital in YTiO$_3$ by synchrotron single-crystal X-ray diffraction is presented.

1. Introduction

  Orbital degeneracy of *d* and *f* valence electrons in strongly correlated materials is an interesting topic in materials science, and observation of the electron density distributions (EDDs) in crystals with large $F(0,0,0)$, which is the number of electrons in a unit cell, is a growing need. Although single-crystal X-ray diffraction (SXRD) is the most straightforward method for determining EDDs, SXRD experiments have technical limitations in accuracy and precision arising from the method of measurement, $\rho = \mathcal{F}^{-1}[F]$, where $\rho$, $\mathcal{F}^{-1}$, and $F$ are the EDD, inverse Fourier transformation operator, and structure factor in complex form, respectively. In the present paper, $\mathcal{F}^{-1}[F]$ is defined as $(1/V_{cell})\int F(\mathbf{r})\exp(-2\pi i\mathbf{hr})d\mathbf{r}$, where $V_{cell}$ is the cell volume in the direct space.

When $\delta\rho$ is defined as the deviation of the observed $\rho$ from the true $\rho$, the first factor determining $\delta\rho$ is the Fourier series truncation at $d_{min} \equiv \lambda/(2\sin\theta_{max})$, and the main effect of this truncation is the uncertainty of $\rho$ inside $\sim d_{min}$ from the atomic positions (see Section 2). The second factor determining $\delta\rho$ is the number of reflections above the noise threshold. If $\delta F$ is the deviation from the true value of $F$, the noise threshold can be expressed as $N \cdot \delta F/F$, for some positive $N$. If errors in $F$ are distributed normally, $N = 3$ would ensure that 99.7% of reflections meeting the criterion $F_{orb}/F > 3\delta F/F$ have some observable contribution from $F_{orb}$. Although the ideal distribution of error for detection of radiations obeys the Poisson distribution, as will be presented in Section 2.3.1 and Appendix H, normal distribution in $F^1$-space (a space with dimension of $F^n$ is called as $F^n$-space throughout the present paper) can provide a better approximation of Poisson distribution than that in $F^2$-space. Therefore, $N = 3$ is considerably a good number in $F^1$-space.

The second factor is significant for most strongly correlated materials with large $F(0,0,0)$. $F_{orb}/F$ generally decreases with increasing $F(0,0,0)$. Thus, the number of reflections satisfying $F_{orb}/F > N \cdot \delta F/F$ decreases without $\delta F/F$ reduction. The history of SXRD technically corresponds to the improvement of $\delta F/F$.

YTiO$_3$, a typical strongly correlated material, is a Mott-insulator and a ferromagnet ($T_C \sim 30$ K) (Greedan., 1985), and it is known to have an ordered arrangement of localized Ti-3$d^1$ electron orbitals (Itho *et al.*, 1999; Ichikawa *et al.*, 2000; Nakao *et al.*, 2002). While the first attempt to determine the arrangement of Ti-3$d^1$ orbitals was performed by SXRD at 298 K and 127 K (Hester *et al.* 1997), early successes determining the arrangement were reported by NMR (Itho *et al.*, 1999), polarized neutron diffraction (PND) (Ichikawa *et al.*, 2000), and X-ray resonant scattering (Nakao *et al.*, 2002). In recent years, successes using synchrotron single-crystal X-ray diffraction (SSXRD), such as SSXRD + PND at 20 K (Voufack *et al.*, 2019; Kibalin *et al.*, 2021), and SSXRD at 25 K (Kitou *et al.*,

2020) have begun to be reported. However, the number of reflections satisfying $F_{orb}/F > N \cdot \delta F/F$ needs to be increased to resolve the following ambiguities. Reports employing SSXRD + PND (Voufack *et al*., 2019; Kibalin *et al*., 2021) explain most of the anisotropic EDDs in the region less than ~0.4 Å from the atomic position of Ti as the anharmonic motion of the Ti atom. Another report using SSXRD (Kitou *et al*., 2020) explains the anisotropy in the same region as the orbital anisotropy of Ti-$3d^1$ orbitals. Although the reference to Ti-$3d^1$ orbital scattering factors in all of these studies (Voufack *et al*., 2019; Kibalin *et al*., 2021; Kitou *et al*., 2020) is restricted to the spherical $\langle j_0 \rangle$, as will be presented in Section 2.1, the aspherical orbital scattering factors $\langle j_2 \rangle$ and $\langle j_4 \rangle$ for Ti-$3d^1$ orbitals make the maximum contribution to $\rho$ at ~0.8 Å and ~0.5 Å in $d$ ($\equiv 1/(2\sin\theta/\lambda)$), respectively (see Figs. 1(b), 2(c) and 2(f)). Thus, the region less than ~0.4 Å from the nucleus should exhibit orbital anisotropy near the nucleus owing to nodal planes passing through the atomic position. However, if strong anharmonic motion of the Ti atom is also present, separately determining the orbital anisotropy and the anharmonic motion becomes difficult (see Section 2.1.1). In cases of large atomic displacement parameters (ADPs), separately determining the orbital anisotropy is better accomplished in the outer region greater than ~0.4 Å from the atomic position, where the effect of the convolution of the probability density function (PDF) for Ti is greatly reduced in magnitude compared to the inner region, and the concavities and convexities of the orbital anisotropy are less smeared by the convoluted PDF (see Fig. 2). For the outer region, the studies employing SSXRD + PND (Voufack *et al*., 2019; Kibalin *et al*., 2021) have provided no EDD synthesized before or after refinement using an anharmonic PDF that included reflections satisfying $d > $ ~0.4 Å. In another study using SSXRD (Kitou *et al*., 2020), no significant anisotropy was detected in the outer region greater than ~0.4 Å. The absence of anisotropy for $d > $ ~0.4 Å suggests the need to improve $\delta F/F$ for satisfying $F_{orb}/F > N \cdot \delta F/F$ for many reflections. This should also clarify the inner $\rho$ region. The present paper focuses on how to reduce $\delta F/F$ and asymptotically approach the accessible limit of EDDs in the theories of SXRD. The present paper first reviews the causes of $\delta F$ observed in most EDD measurements and aspherical orbital scattering factors such as $\langle j_k \rangle$. Then, methods are proposed for reducing $\delta F$ theoretically and experimentally through observation of Ti-$3d^1$ orbitals in YTiO$_3$ by SSXRD. Since

$d_{min}$ in the present study is ~0.42 Å, the targeted regions are those with anisotropies greater than ~0.4 Å (a bit smaller than $d_{min}$ for high-accuracy data) from the nucleus.

The present paper denotes complex numbers or vectors in linear algebra with bold type fonts and denotes scalars such as the modulus of complex numbers with non-bold type fonts. $|F|$ and $F$ represent the identical quantity.

## 2. Theoretical

Reducing $\delta\rho$ to ~0 is the technical goal in SXRD. Before exploring the details of this topic, the basic strategy for reducing $\delta\rho$ is presented.

The superposition of waves $\rho = \mathcal{F}^{-1}[F]$ defines the possibilities and limitations of SXRD. The first limitation in a real experiment is $\delta\rho$ due to the Fourier series truncation error, which cuts off waves at $d_{min}$ ($\equiv 1/(2\sin\theta_{max}/\lambda)$). The typical effects of the truncation error are (A) the ambiguity of $\rho$ within ~$d_{min}$ from the nuclear position and (B) the slight roughness in $\rho$ due to truncation of waves shorter than $d_{min}$. The reason for (A) is that the actual local origin of the summation of waves is the atomic position, and waves shorter than $d_{min}$ are not superimposed. (Recall the equation $F(\mathbf{h}) = \sum_{atom} f_{atom}(\mathbf{h})\exp(2\pi i \mathbf{h}\mathbf{r}_{atom})$, and put the equation into $\rho(\mathbf{r}) = (1/V_{cell})\sum_{\mathbf{h}} F(\mathbf{h})\exp(-2\pi i \mathbf{h}\mathbf{r})$, where $f_{atom}$, $\mathbf{r}_{atom}$, and $V_{cell}$ are the atomic scattering factor, atomic position, and cell volume, respectively. Then, $\rho(\mathbf{r}) = (1/V_{cell})\sum_{atom}\sum_{\mathbf{h}} f_{atom}(\mathbf{h})\exp\{-2\pi i \mathbf{h}(\mathbf{r}-\mathbf{r}_{atom})\}$ is obtained, and one can see that the inverse Fourier summation for each atom, *i.e.*, $\rho_{atom}(\mathbf{r}) = \sum_{\mathbf{h}} f_{atom}(\mathbf{h})\exp\{-2\pi i \mathbf{h}(\mathbf{r}-\mathbf{r}_{atom})\}$, is summed over the atoms in the equation of $\rho(\mathbf{r})$.)

The second limitation in the experimental $\rho$ comes from the Poisson statistics, which the detection of radiation obeys and is an unremovable cause of $\delta F$. Although the method of measurement $\rho(\mathbf{r}) = (1/V_{cell})\sum_{\mathbf{h}} F(\mathbf{h})\exp(-2\pi i \mathbf{h}\mathbf{r})$ indicates that each accurate $F(\mathbf{h})$ should be summed up with uniform weight, an accurate $F$ with no error is unattainable. Thus, general statistical refinements employ weights such as $(1/\delta F)^n$ for each reflections ($n = 1$ for refinements on $F$, and $n = 2$ for refinements on $F^2$ or $I$). Two keys for achieving the theoretically observable limit of $\rho$ with only the unavoidable truncation error are: (A) the reduction of $\delta F/F$ to ~0, and (B) the realization of uniform weighting under Poisson statistics.

Topics on $F_{orb}/F$ (Section 2.1), $\delta F/F$ in data collection (Section 2.2), uniform weighting under Poisson statistics (Section 2.2), and $\delta F/F$ in data refinement (Section 2.3) are treated in the present section.

## 2.1. Required noise-to-signal ratio $F_{orb}/F$ to detect targeted orbital

How small a $\delta F/F$ is required for observation of the Ti-$3d^1$ orbital in YTiO$_3$ depends on the magnitude of $F_{orb}/F$. The $F_{orb}/F$ for Ti-$3d^1$ orbitals in YTiO$_3$ is reviewed below.

In the following calculation of orbital scattering factors $\langle j_k \rangle$, the direct Fourier transformation of $\langle \varphi | \varphi \rangle$ introduced by Stewart (1969) is used. Here, $|\varphi\rangle$ is an orbital wavefunction of which the angular part is a spherical harmonic function. The formalism provides the foundation for X-ray atomic orbital (XAO) analysis (Tanaka *et al.*, 2008; Tanaka & Takenaka, 2012). The advantage of the formalism compared to multipole formalisms including the one proposed by Hansen & Coppens (1978) is that the direct Fourier transformation of $\langle \varphi | \varphi \rangle$ leads to no error in $\langle j_k \rangle$ calculation. In multipole formalisms, decomposition of $\langle \varphi | \varphi \rangle$ into multipoles requires numerical curve fitting by, for example, the least-squares method and errors depending on the refinement models and sampling methods are unavoidable.

### 2.1.1. Aspherical components of orbital scattering factors

The formalism of the direct Fourier transformation of $\langle \varphi | \varphi \rangle$ is summarized in Appendix A. For the case of a $d$-orbital with the azimuthal quantum number $l = 2$, nonzero $\langle j_k \rangle$ are available only for $k = 0$, 2, and 4. Figure 1 shows $\langle j_0 \rangle$, $\langle j_2 \rangle$, and $\langle j_4 \rangle$ for the Ti-$3d^1$ orbital in YTiO$_3$. The radial wavefunction used for the calculation is constructed by the linear combination of Slater-type orbitals (STOs) computed by Clementi & Roetti (1974) for a neutral Ti atom. Figure 1(a) describes $\langle j_0 \rangle$, $\langle j_2 \rangle$, and $\langle j_4 \rangle$ as a function of $\sin\theta/\lambda$, and Fig. 1(b) describes $4\pi V^{2/3}_{cell} (2\sin\theta/\lambda)^2 \langle j_k \rangle$ for $k = 0, 2,$ and 4. The multiplicative term $4\pi V^{2/3}_{cell} (2\sin\theta/\lambda)^2$ for $\langle j_k \rangle$ in Fig. 1(b) corresponds to the number of reflections at $\sin\theta/\lambda$ for YTiO$_3$. ($4\pi V^{2/3}_{cell} (2\sin\theta/\lambda)^2$ is the number of reflections on a sphere whose radius is $V^{1/3}_{cell} (2\sin\theta/\lambda)$.) Therefore, Fig. 1(b) exhibits a more realistic contribution from $\langle j_k \rangle$ to $\rho$ at $\sin\theta/\lambda$.

However, $\langle j_k \rangle$ are only the radial scattering factors, and contributions to $\rho$ from $\langle j_2 \rangle$ and $\langle j_4 \rangle$ vanish when the angular part is spherical (see Appendix A). For seeing the realistic contribution rate from an aspherical orbital to $\rho$, a Fourier synthesis reflecting all the coefficients in equation (A4) is required, and Fig. 2 shows this using the coefficients determined from the present study. To see the effect of the truncation error, Fig. 2 is synthesized for different $\sin\theta/\lambda$ ranges. From the left, the $\sin\theta/\lambda$ ranges are 0.0 to 2.0 Å$^{-1}$, 0.0 to 1.2 Å$^{-1}$, and 1.2 to 2.0 Å$^{-1}$. To see another effect of the thermal smearing of $\rho$ caused by convolution of the Debye-Waller factor, the harmonic vibrational model of the PDF determined at 298 K for the Ti atom in the present study is convoluted in the bottom panels of Figs. 2(d)-2(f). Owing to the nature of the superposition of waves, the following equations are satisfied for $\rho$: Fig. 2(a) = Fig. 2(b) + Fig. 2(c), and Fig. 2(d) = Fig. 2(e) + Fig. 2(f).

The conclusion of Figs. 1 and 2 is the following. (A) The maximum contribution rates to $\rho$ from $\langle j_2 \rangle$ and $\langle j_4 \rangle$ in $d$ are ~0.8 and ~0.5 Å, respectively (~0.63 and ~1.0 Å$^{-1}$ in $\sin\theta/\lambda$) (see Fig. 1(b)), and $\langle j_2 \rangle$ and $\langle j_4 \rangle$ exhibit larger contributions to $\rho$ for higher-angle reflections compared to the spherical $\langle j_0 \rangle$, whose maximum contribution to $\rho$ in $d$ is ~2.0 Å (~0.25 Å$^{-1}$ in $\sin\theta/\lambda$). (B) The effect of thermal motion is moderate far away from the nucleus; whereas the area satisfying $\rho(\mathbf{r}) = 0$ near the nucleus ($d < \sim 0.4$ Å) corresponding to nodal planes observed in Fig. 2(a) disappeared in Fig. 2(d), $\rho(\mathbf{r})$ at larger $d$ ($> \sim 0.4$ Å) from the nucleus exhibits almost no change by the convolution of the Debye-Waller factor. (Although $3\sigma(\mathbf{u})$ in the PDF for Ti1 is less than 0.2 Å (see $U_{ij}$ in Table. 2), the spatial scale of concavities and convexities of the Ti-3$d^1$ orbital is much larger than 0.2 Å at $d > \sim 0.4$ Å from the nucleus. Therefore, the convoluted PDF behaves like a $\delta$ function (note that $\delta*f = f$, where * is the symbol of convolution and $f$ is a function). Conversely, if the convoluted two functions of PDF and static $\rho$ have comparative spatial scales for concavities and convexities and have unknown shape, restoring the PDF and the static $\rho$ from the observed (dynamic) $\rho$ generally become difficult.) (C)

When a certain region of $d$ is truncated, the synthesized $\rho$ exhibits almost no anisotropy in the truncated region of $d$ from the nucleus, since the waves which can describe concavities and convexities at a certain $d$ from the nucleus are not provided in the equation $\rho_{atom}(\mathbf{r}) = \sum_{\mathbf{h}} f_{atom}(\mathbf{h})\exp\{-2\pi i\mathbf{h}(\mathbf{r}-\mathbf{r}_{atom})\}$. (Note that this problem is not specific for Fourier synthesis, since the truncation of waves at a certain $d$ from the nucleus affects the result of the refinement itself. In maximum entropy method (MEM), another factor what phase is attached to $\mathbf{F}$ also affects the resulting EDDs. Brief discussions on this point are summarized in Appendix J.) In terms of the present context discussing $\rho_{atom}(\mathbf{r}) = \sum_{\mathbf{h}} f_{atom}(\mathbf{h})\exp\{-2\pi i\mathbf{h}(\mathbf{r}-\mathbf{r}_{atom})\}$, reducing weight for reflections with large uncertainty causes similar effect to truncation, since removal of some reflections and reduce of weight for some reflections has no significant difference when the weight is greatly reduced. In Supporting Information, $\rho$ synthesized with a cut-off at a certain threshold of modulus of $F$ are shown (Figs. S5 and S6). Since, each independent reflections corresponds to each independent waves in the direct space, the lost information by reducing the weight or truncation or rejection cannot be compensated by any other reflections.

### 2.1.2. $F_{orb}/F$ for Ti-$3d^1$ orbitals in YTiO$_3$

$F_{orb}/F$ are constants under a given set of experimental conditions, and how small a $\delta F/F$ is required to satisfy $F_{orb}/F > \delta F/F$ depends on the magnitude of $F_{orb}/F$. Before comparing the magnitudes of $F_{orb}/F$ and $\delta F/F$, which is presented in Section 6.2, $F_{orb}/F$ for Ti-$3d^1$ orbitals in YTiO$_3$ is reviewed.

$F_{orb}/F$ can be defined directly with $|F_{orb}|/|F|$. However, the change in $|F|$ due to the contribution of $\boldsymbol{F}_{orb}$ is experimentally observed as the projected length of $\boldsymbol{F}_{orb}$ on a unit vector $\boldsymbol{F}/|F|$ with a plus (minus) sign for increase (decrease). Therefore, the term "signed $F_{orb}/F$," defined as the proportion of ($\boldsymbol{F}_{orb} \cdot \boldsymbol{F}/|F|$) to $|F|$, is introduced and used as a practical measure of $F_{orb}/F$ throughout the present paper:

$$\text{'Signed } F_{orb}/F\text{'} = (\boldsymbol{F}_{orb} \cdot \boldsymbol{F})/|F|^2. \tag{1}$$

Figure 3(a) shows $(\boldsymbol{F}^0_{\mathrm{orb}} \cdot \boldsymbol{F}^0)/|\boldsymbol{F}^0|^2$ versus $\sin\theta/\lambda$ for the Ti-$3d^1$ orbital in YTiO$_3$, where the superscript 0 denotes exclusion of the anomalous dispersion terms. It can be seen that only a small proportion of reflections will have large $(\boldsymbol{F}^0_{\mathrm{orb}} \cdot \boldsymbol{F}^0)/|\boldsymbol{F}^0|^2$. Whereas the signal from Ti-$3d^1$ orbitals is almost unobservable for reflections with almost zero $(\boldsymbol{F}^0_{\mathrm{orb}} \cdot \boldsymbol{F}^0)/|\boldsymbol{F}^0|^2$ magnitude, reflections with relatively large magnitudes of $(\boldsymbol{F}^0_{\mathrm{orb}} \cdot \boldsymbol{F}^0)/|\boldsymbol{F}^0|^2$ should be collected with the condition $(\boldsymbol{F}^0_{\mathrm{orb}} \cdot \boldsymbol{F}^0)/|\boldsymbol{F}^0|^2 > \delta F/F$ for observation of Ti-$3d^1$ orbitals. Figure 3(b) shows $(\boldsymbol{F}_{\mathrm{orb}} \cdot \boldsymbol{F})/|\boldsymbol{F}|^2$ with anomalous dispersion terms for a wavelength of 0.75 Å. Although $\sin\theta/\lambda$ values greater than 1.333 Å$^{-1}$ are inaccessible by X-rays with a wavelength of 0.75 Å, the region outside the limiting sphere provides a simulation for shorter wavelengths. The decrease of $(\boldsymbol{F}_{\mathrm{orb}} \cdot \boldsymbol{F})/|\boldsymbol{F}|^2$ by inclusion of anomalous dispersion terms exhibited in Fig. 3(b). As presented in Appendix B, although small $F$ is significant to increase the magnitude of $(\boldsymbol{F}_{\mathrm{orb}} \cdot \boldsymbol{F})/|\boldsymbol{F}|^2$, imaginary parts of anomalous dispersion terms generally prevent reflections from becoming $F \approx 0$. Therefore, choosing smaller imaginary parts due to anomalous dispersions is preferable, when the resolution $d_{\mathrm{min}}$ is not severely restricted by the choice of X-ray energy.

**2.2. Lower limit of noise-to-signal ratio $\delta F/F$ in data collection**

The present subsection focuses on $\delta F/F$ in data collection. Section 2.2.1 reviews the consequences of Poisson statistics in $\delta F/F$, namely, that the application of uniform weighting is approximately realizable and the satisfaction of $\delta F/F \approx 0$ is also possible. Section 2.2.2 introduces the additional contribution to $\delta F/F$ due to the accumulation of errors originating from the data collection equipment. As the $\delta F/F$ component from Poisson statistics approaches ~0 at large counts, the contribution from the data collection equipment to $\delta F/F$ becomes the dominant factor defining the lower limit of $\delta F/F$ (Miyahara *et al.*, 1986; Amemiya & Chikawa, 2006). Section 2.2.2 also introduces a simple model expressing $\delta F/F$ as the sum of the $\delta F/F$ due to Poisson statistics and the $\delta F/F$ due to the data collection and reduction procedure. Section 2.2.3 discusses the $\delta F/F$ contributed by multiple diffraction (MD), which changes $\delta F/F$ depending on the relative orientation between the incident beam and the crystal.

### 2.2.1. $\delta F/F$ contributed by Poisson statistics

The detection of radiation obeys Poisson statistics. Although a $\delta F/F$ contributed by Poisson statistics is unavoidable, the $\delta F/F$ is reducible and an asymptotic approach to $\delta F/F = 0$ is possible.

Let $I_{count}$ and $F_{count}$ be, respectively, the count collected by a detector and its square root $(I_{count})^{1/2}$. Then, Poisson statistics derives $\sigma(F_{count})$ as 0.5 (see Appendix C). If the scaling factor $k^0_{sc}$ satisfying $F_{count} = k^0_{sc} F$ is introduced, $\sigma(F_{count}) = 0.5$ is transformed into $\sigma(F) = 0.5/k^0_{sc}$. For an infinite number of trials, $\delta F/F$ converges to $\sigma(F)/F$, i.e., $0.5/(k^0_{sc} F)$. Thus, $k^0_{sc}$ is a unique parameter available for $\delta F/F$ reduction. $k^0_{sc}$ is defined by (see eq. D2)

$$k^0_{sc} = k_{sc} (Lp\, A\, Y\, O)^{1/2}, \qquad (2)$$

where $k_{sc}$, $Lp$, $A$, $Y$, and $O$, are, respectively, the scale factor, Lorentz-polarization factor, absorption factor, extinction factor, and sum of all non-corrected factors. Since $O$ includes a model accounting for the errors generated in every trial whose mechanisms are not elucidated yet, the error of $k^0_{sc}$ is not treated explicitly throughout this paper. As described in Appendix D, although the dimensions of factors constituting $k^0_{sc}$ varies depending on the experimental conditions, enlargement of $r$, $I_0$, and $t$ is effective for increasing $k^0_{sc}$, where $r$ is the crystal radius for spherical crystals, $I_0$ is the incident beam intensity, and $t$ is the intensity collection time.

### 2.2.2. Saturation of $\delta F/F$ in the data collection equipment

An additional contribution to $\delta F/F$ is the accumulation of errors in the data collection equipment, which defines the lower limit of $\delta F/F$ at relatively large $F_{count}$, where $\delta F/F$ due to Poisson statistics approaches ~0 (Miyahara et al., 1986; Amemiya & Chikawa, 2006). Evaluation of the sum of $\delta F/F$ from Poisson statistics and that from the data collection equipment is necessary for the diagnosis of $\delta F/F$ whose constituent factors are not elucidated yet. The evaluated $\delta F/F$ is also necessary for the application of suitable weights $(1/\delta F_{estimated})^2$ in statistical refinements such as the least-squares method, for which the minimization function is defined as $\Sigma((F_{obs} - F_{calc})^2/(\delta F_{estimated})^2)$. Since the true magnitude of $\delta F$ cannot be determined, the true weight $(1/\delta F)^2$ is replaced with $(1/\delta F_{estimated})^2$, where $\delta F_{estimated}$ is the estimated $\delta F$. For diagnosis of the error hiding in the data, each $\delta F$ is better to be treated without averaging. Therefore, the weight is expressed as $(1/\delta F_{estimated})^2$ instead of $(1/\sigma(F))^2$ throughout the present paper.

Let $I_{count}$ and $I_{corr}$ be, respectively, the output counts by the system and the counts internally detected by the system. In $I_{corr}$, system-specific corrections such as background subtraction are applied, and the following equation connecting $I_{count}$ and $I_{corr}$ is introduced (see eq. F1):

$$I_{count} = I_{corr} M_{sc}, \qquad (3)$$

where $M_{sc}$ is a scaling factor converting $I_{corr}$ to $I_{count}$. By the derivation in Appendix F, equation (3) leads to $\sigma(F_{obs})$ as (see eq. F5):

$$\sigma(F_{obs}) = \left((\sigma'(F_{obs}))^2 + \left(\frac{\sigma(M_{sc})}{2\,M_{sc}}\right)^2 F_{obs}^2\right)^{\frac{1}{2}}, \qquad (4a)$$

where (see eq. F4)

$$\sigma'(F_{obs}) = \frac{M_{sc}\,\sigma(I_{corr})}{2\,k_{sc}^0\,F_{count}}. \qquad (4b)$$

Although $\sigma(I_{corr})$ in equation (4b) depends on background counts and other system-specific parameters, $\sigma(I_{corr})$ is generally expected to approach $(I_{corr})^{1/2}$ at large $I_{corr}$ since the detection of radiation basically obeys Poisson statistics. Therefore, the right-hand side of equation (4b) approaches $0.5/k^0_{sc}$ at large counts if $M_{sc} \approx 1$ and $M_{sc}$ is independent from $I_{corr}$. Then, the following equation is derived (for the case of $M_{sc} \neq 1$, $k^0_{sc}$ in the following equations is replaced with $k^0_{sc}/M_{sc}^{1/2}$):

$$\sigma(F_{obs}) \approx \left(\left(\frac{0.5}{k_{sc}^0}\right)^2 + \left(\frac{\sigma(M_{sc})}{2\,M_{sc}}\right)^2 F_{obs}^2\right)^{\frac{1}{2}}. \qquad (5)$$

Division of equation (5) by $F_{obs}$ gives $\sigma(F_{obs})/(F_{obs})$, *i.e.*, $\sigma(I_{obs})/(2\,I_{obs})$, as:

$$\frac{\sigma(F_{obs})}{F_{obs}} = \frac{\sigma(I_{obs})}{2\,I_{obs}} \approx \left(\left(\frac{0.5}{k_{sc}^0\,F_{obs}}\right)^2 + \left(\frac{\sigma(M_{sc})}{2\,M_{sc}}\right)^2\right)^{\frac{1}{2}}. \qquad (6)$$

Equation (6) explains the cause of saturation of $\sigma(F_{obs})/F_{obs}$ at relatively large counts as $\sigma(M_{sc})/(2M_{sc})$, since $0.5/(k^0_{sc} F_{obs})$ is ~0 at large counts.

According to the definition of $M_{sc}$, one can predict that detectors having sensitivity nonuniformity and its precise correction for each pixel is not applied or accurate correction is not possible due to dead-area causing counting-loss in the detection area will give rise to large $\sigma(M_{sc})$. The early PILATUS 1M has amounting to 7.5% of dead-area locating at the corners of pixels, where the total charges generated by a photon are shared among the adjacent pixels and each of the pulse-heights is mostly lower than the thresholds for detection (Broennimann *et al.*, 2006). Since each of the diffraction spots observed in general SXRD measurements spread over several pixels, the counting-loss at the pixel corners is crucial for accurate integrated intensity collections (Shanks, 2014). Therefore, the PILATUS detector (Broennimann *et al.*, 2006) should have large $\sigma(M_{sc})/(2M_{sc})$ in practical applications of SXRD. As a remedy for counting-loss due to charge sharing, integrating an inter-pixel communication functionality in the readout ASIC (application specific integrated circuit) (Ballabriga, R. *et al.*, 2007; Gimenez, E. N. *et al.*, 2011) is developed and available in some detector families, *e.g.*, Medipix3 (R. Ballabriga, *et al.*, 2007; Gimenez, E. N. *et al.*, 2011). However, this approach suppresses the count rate capability (Forjdh, E. *et al.*, 2014). In imaging plate (IP) detectors, $3\delta F/F$ is reported to be 2-3% (Miyahara *et al.*, 1986; Amemiya & Chikawa, 2006). In zero-dimensional detectors, which always receive all the diffraction spots at the center of each detection device, $\sigma(M_{sc})/(2M_{sc})$ is mostly negligible. Thus, the dominant factor determining the lower bound of $\delta F/F$ for zero-dimensional detectors in a one-second measurement is $(I_{saturate})^{-1/2}$, where $I_{saturate}$ is the upper boundary of cps (counts per second) holding count rate linearity. The stacked avalanche photodiode (APD) detector, which stacks several APD devices along the incoming X-ray direction (Kishimoto, 1998; Kishimoto *et al.*, 1998; Kishimoto & Seto, 2007), has an $I_{saturate}$ of ~$10^8$ cps, and is one of the most appropriate detectors for EDD measurements by SSXRD. (Even for recent two-dimensional single-photon counting detectors such as PILATUS 3X CdTe (DETECTRIS) and EIGER2 (DETEECTRIS), their dynamic range are below ~$10^7$cps/pixel even when correction of count-rate is applied (Krause *et al.*, 2020; Donath *et al.*, 2023).)

The rest of the present subsubsection introduces a simple equation that is convenient for plotting $\delta F/F$ versus $F$. The equation can also be used to determine the rejection criterion for outliers. On replacing $k^0_{sc}$ with $k_{sc}$ and applying the inequality $(a + b) \geq (a^2 + b^2)^{1/2}$, which is satisfied for non-negative $a$ and $b$, equation (6) is reduced to the following equation:

$$\frac{\delta F}{F} \approx \frac{0.5}{k_{sc} F_{obs}} + \frac{\sigma(M_{sc})}{2 M_{sc}}. \qquad (7)$$

Although replacing $k^0_{sc}$ with a constant number $k_{sc}$ is a bit of a rough approximation, this removes parameters causing $\delta F/F$ to be unplottable against $F$. The application of the inequality $(a + b) \geq (a^2 + b^2)^{1/2}$ loosens the rejection criterion and is preferable in practical uses. (In general data collections, as $F_{obs}$ decreases, $\sigma(F_{obs})$ become larger than $0.5/(k_{sc}F_{obs})$, since an error propagated from background count which is proportional to $(I_{background})^{1/2}$ become comparative to $(I_{count})^{1/2}$ in magnitude. Therefore, application of the inequality may provide a slightly better approximation of $\delta F/F$ practically. Thus, adopting the symbol $\approx$ in equation (7) is permissible.)

By defining $\langle F_{obs} \rangle$ as the symmetry equivalents, a simultaneous plot of equation (7) and $|F_{obs} - \langle F_{obs} \rangle|/\langle F_{obs} \rangle$ versus $\langle F_{obs} \rangle$ can provide visual information showing which reflections have $|F_{obs} - \langle F_{obs} \rangle|/\langle F_{obs} \rangle$ that is too large compared to $N \cdot \delta F/F$ for the data collection equipment.

### 2.2.3. $\delta F/F$ from multiple diffraction

$\delta F/F$ from multiple diffraction (MD) is systematic and cannot be treated statistically by the $\delta F/F$ model described in Section 2.2.2, and so, correction or avoidance is necessary. Since the effect of MD on the $R(F)$-factor ($\equiv \Sigma|F_{obs} - F_{calc}|/\Sigma F_{obs}$) is generally limited, datasets even with no avoidance or no correction for MD generally yield $R(F)$-factors smaller than 2%. However, MD significantly decreases the number of reflections satisfying $F_{orb}/F > N \cdot \delta F/F$, especially for small-$F$ reflections.

MDs are the re-diffractions among the reflections lying simultaneously on the Ewald sphere, and cause a loss and gain of intensity. The loss of intensity is known as the *Aufhellung* process (Wagner, 1920), and the gain in intensity is known as the *Umweganregung* process (Renninger, 1937). The

*Umweganregung* process is the chief cause for the significant decrease in the number of reflections satisfying $F_{orb}/F > N \cdot \delta F/F$ for reflections with small $F$. The reason can be explained as follows. Let $\boldsymbol{F}_{MD}$ be the $\delta F$ caused by MD. If all the other causes of $\delta F$ are omitted, the equation $\boldsymbol{F}_{obs} = \boldsymbol{F} + \boldsymbol{F}_{MD}$ is satisfied. As the true magnitude of $|\boldsymbol{F}|$ decreases and $|\boldsymbol{F}_{MD}| > 2|\boldsymbol{F}|$ becomes satisfied, regardless of the relative angle between $\boldsymbol{F}_{MD}$ and $\boldsymbol{F}$, a gain in intensity, *i.e.*, $|\boldsymbol{F}_{obs}| > |\boldsymbol{F}|$, occurs (Let $\boldsymbol{F}$ and $\boldsymbol{F}_{MD}$ be defined as $\boldsymbol{F} = |\boldsymbol{F}|\exp(i\alpha)$ and $\boldsymbol{F}_{MD} = 2|\boldsymbol{F}|\exp(i\beta)$. Then, $\boldsymbol{F}_{obs} = \boldsymbol{F}\{1 + 2\exp\{i(\beta - \alpha)\}\}$ is derived, and $|\boldsymbol{F}_{obs}|$ varies from $|\boldsymbol{F}|$ to $3|\boldsymbol{F}|$. Therefore, $|\boldsymbol{F}_{obs}| \geq |\boldsymbol{F}|$ is constantly satisfied). As $|\boldsymbol{F}|$ decreases, the likelihood of a gain in intensity increases.

If the reliability factor $\delta F/F$ is evaluated by $|\boldsymbol{F}_{MD}|/|\boldsymbol{F}|$, the condition $|\boldsymbol{F}_{MD}| > 2|\boldsymbol{F}|$ gives $\delta F/F > 200\%$. In general, the observed $F$ with $\delta F/F > 50\%$ contains no reliable information on the true $\rho$, and adds noise to $\rho$. This is one of the chief reasons why most experimental $\rho$ cannot be observed as clearly as the $\rho$ shown in Fig. 2. Since large $F_{orb}/F$ in crystals with large $F(0,0,0)$ is available only for reflections with small $F$ (will be shown in Fig.10, and note that $(\boldsymbol{F}_{orb} \cdot \boldsymbol{F})/|\boldsymbol{F}|^2$ become large only for small $F$), the reliability factor $\delta F/F$ is generally enlarged by MD for small-$F$ reflections.

As well known, superlattice reflections have critical importance in determination of a supercell structure. Although superlattice reflections and small-$F$ reflections are common in having small-$F$, superlattice reflections are free from re-diffraction from the fundamental reflections and have very small $\delta F/F$; since, no route of re-diffraction at fractional Miller indices of a superlattice structure is constructed only by the integer Miller indices of the fundamental structure. If the $\delta F/F$ from MD is reduced in fundamental small-$F$ reflections, the detailed orbital anisotropies in $\rho$ should become clear.

### 2.3. $\delta F/F$ in data refinements

Errors in the refinement models and the weightings for reflections can be additional causes of $\delta F/F$. In the present subsection, $\delta F/F$ in refinements are discussed.

### 2.3.1. Effect of approximation error in refinements on $F^2$ or $I$

Whether the $F^2$-space quantities of $I$ and $F^2$ or the $F^1$-space quantity of $F$ should be used as the input for the refinement has been a matter of debate (Schwarzenbach *et al.*, 1989). However, using $F$ for the refinement of $\rho$ can remove an approximation error contained in refinements on $F^2$-space quantities. Since the method of measurement, *i.e.*, $F = \mathcal{F}[\rho]$, is defined in the space of $F^1$, no crystal structure analysis can avoid the process of converting $F^n$-space-based quantities to $F^1$-space-based quantities in the refinement. If the conversion from $F^n$-space-based quantities to $F^1$-space-based quantities is implemented in non-linear optimization software programs, the conversion error approximating $F^1$-space-based quantities with a derivation of $F^n$-space-based quantities is generated and prevents convergence. Appendix G treats this topic, and the following is a summary of the discussion.

The magnitude of $\Delta F$, which is required in each refinement cycle, is approximated by $\Delta F^n/(\partial(F^n)/\partial F)$ when the refinement is executed in $F^n$-space. A graphical representation is shown in Fig. 4. $\partial(F^n)/\partial F$ is the slope at the point $(F, F^n)$. One can see that the difference between $\Delta F$ and $\Delta F^n/(\partial(F^n)/\partial F)$ grows when $\Delta F^n$ is large or $n$ is far from 1. Although the difference between $\Delta F$ and $\Delta F^n/(\partial(F^n)/\partial F)$ becomes small as $\Delta F^n$ approaches zero, the actual experimental data with errors never satisfy the condition $\Delta F^n = 0$. Thus, the systematic error persists even when convergence is reached, and the parameters optimized by the refinements on $F^n$ (other than $n = 1$) are still optimizable by refinements on $F$.

Another bias found in the refinement on $F^2$ is the lower precision in approximating the Poisson distribution with a normal distribution in $F^2$-space as compared to $F^1$-space. Since the least-squares method is a special case of the maximum likelihood method when the distribution function is a normal distribution (Prince & Collins, 2006), the error distribution model of the datasets refined by least-squares method should not differ greatly from the normal distribution. Although a previous work (Wilson, 1979) drew the opposite conclusion, *i.e.*, that $F^1$-space quantities are distorted by non-linear projection of the square root from $F^2$-space quantities, from what will be described in the present subsubsection, that study omits the process of placing a certain distribution function in $F^2$-space and projecting the function on $F^1$-space. Thus, practical quantitative evaluations were not actually performed.

As presented above, our goal is to remove all $\delta F$ other than the unremovable contribution from Poisson statistics. The projection of the Poisson distribution from $F^2$-space to $F^1$-space is discussed in Appendix H. The graphical representations of the projected Poisson distribution and its approximation by the normal distribution in, respectively, $F^1$- and $F^2$-space for $I_{count} = 4$, *i.e.*, $F_{count} = 2$, are shown in Figs. 4(b) and 4(c). The dots in Figs. 4(b) and 4(c) are the Poisson distributions, and the dashed lines are the approximated normal distributions with $\sigma(F^2_{count}) = 2$ for Fig. 4(b) and $\sigma(F_{count}) = 0.5$ for Fig. 4(c). As Figs. 4(b) and 4(c) show, whereas the probability of the normal distribution at $I_{count} = 0$ in $F^2$-space is still large, the normal distribution in $F^1$-space almost converges to 0 at $F_{count} = 0$. Thus, the normal distribution holds better in $F^1$-space than in $F^2$-space. Since $3\sigma(F_{count})$ (= 1.5) and $\sqrt{2}$ are close in magnitude, the normal distribution in $F^1$-space gives an adequate approximation even for $I_{count} = 2$, *i.e.*, $F_{count} = \sqrt{2}$ (see Fig. S7 in Supporting Information).

This result indicates another point: that the necessity of the maximum likelihood method in refinements for more precision decreases as the various sources of $\delta F$ are removed and the error distribution model converges to a Poisson distribution.

### 2.3.2. Artificial control of weights

The quantity $C^2$ is defined as $\chi^2/(N_{refl} - N_{param})$, where $\chi^2$ is $\Sigma((F_{obs} - F_{calc})/\delta F_{estimated})^2$ and $N_{param}$ is the number of parameters. The goodness-of-fit is a factor defined as the square root of $C^2$, and both the goodness-of-fit and $C^2$ approach unity if all the user-defined models, *i.e.*, $F_{calc}$ and $\delta F_{estimated}$, approach the true models generating the data. However, the true models of $F_{obs}$ and $\delta F$ cannot be obtained. Therefore, the requirement for the goodness-of-fit to approach unity is always violated in real experiments.

As the following quote (Spagna & Camalli, 1999) indicates, artificial control of weights $(1/\delta F_{estimated})^2$ to approach a goodness-of-fit of unity is a temporary approach, and we should surpass this by removing as many sources for $\delta F$ as possible: '*If the value of $S^2$ (= $C^2$ in this paper) approaches unity and W (= $(1/\delta F_{estimated})^2$ in this paper) is valid, then the $C_j$ (= $F_{calc}$ in this paper) are*

*well estimated and the $\sigma_j$ (= $\delta F_{estimated}$ in this paper) are reliable. However, we know that W is usually invalid; it may be postulated that if $S^2$ approaches unity and $C_j$ are valid, then W is possibly valid.*'

As mentioned above, removal of any sources of $\delta F$ other than the Poisson statistics is a goal for the EDD measurements. In many refinements, one of the most common formulae is the weight expressed as $(1/\delta F_{estimated})^2 = 1/\{\sigma^2(F_{obs}) + a F^2_{obs}\}$, where $a$ is an adjustable parameter to make the goodness-of-fit close to unity. This formula is available in most charge density refinement programs. However, as presented in Section 2.2.2, the parameter $a$ corresponds to $\sigma(M_{sc})/(2M_{sc})$. If $\sigma(M_{sc})/(2M_{sc})$ is set to 0.1, the physical meaning is that $3\sigma(F_{obs})$ of $F_{obs}$ is 30% on average. This is unsuitable for data with small $R(F)$ factors, such as $R(F) < 2\%$. Owing to the various sources of $\delta F$ such as MD, relatively large disagreements between $F_{obs}$ and $F_{calc}$ are normal for reflections with small $F$, and satisfaction of the equation $\Sigma(|F_{obs} - F_{calc}|/\delta F_{estimated}) > \sqrt{(N_{refl} - N_{param})}$, *i.e., 'the goodness-of-fit'* > 1, is also normal. Once $\delta F_{estimated}$ is estimated or evaluated as precisely as possible, one should not change $\delta F_{estimated}$ significantly to make the goodness-of-fit approach unity.

As a related topic, another popular weight, $(1/\delta F_{estimated})^2 = 1/\{a + b (1/3F^2_{obs} + 2/3F^2_{calc})\}$ (Wilson, 1976), also needs to be revisited. As summarized in Appendix I, the derivation of the weight violates some requirements in the least-squares method, and the weight prevents the maximum use of the experimental data $F_{obs}$, since $F_{calc}$ to be determined from $F_{obs}$ depends on $F_{calc}$ itself through the weight.

### 3. Experimental

The effects of (A) MD and (B) Poisson statistics are examined through the observation of the Ti-$3d^1$ orbital in YTiO$_3$. Experiment A compares the two datasets one with MD avoidance (MDA), the other which is measured at bisecting positions, non-MDA. All the datasets in Experiment A are collected by an APD detector (Kishimoto & Seto, 2007). Experiment B applies MDA, and two datasets collected by the APD detector and by a scintillation detector (SD) are compared.

### 3.1. Experimental station

The experiments were carried out using the four-circle diffractometer installed at the beamline BL-14A of the Photon Factory, Institute of Materials Science, High Energy Accelerator Research Organization (Satow *et al*., 1989; Vaalsta & Hester, 1997). The three circles for the crystal rotation in four-circle diffractometers enable $\psi$-rotation using the method described by Busing & Levy (1967), and the $\psi$-rotation is necessary for MDA. Figure 5 shows a schematic view of the beamline, drawn by *OpenSCAD* (Marius Kintel). Owing to the vertically polarized beam generated by a vertical wiggler, the scanning direction of the BL-14A is in a horizontal plane. Thus, high-speed positioning of the goniometer and a stable scan rate are achieved. The monochromatic X-rays were generated by a double-crystal Si-111 monochromator, and the beam was focused by a curved fused-quartz mirror coated with rhodium. The unwanted harmonics were reduced by the focusing. The intensity fluctuation of the incident beam was monitored by an ion-chamber and used for correction.

### 3.2. Specimen

A single crystal of $YTiO_3$ was prepared by the floating-zone method. The grown crystal was crushed and rounded into a sphere 132 μm in diameter (see Fig. S8) by the Bond method (Bond, 1951) with minor modifications.

Figure 6 shows the crystal structure of $YTiO_3$, as drawn by *VESTA* (Momma & Izumi, 2011). $YTiO_3$ belongs to the space group *Pnma*, and the lattice parameters determined by the present study are $a$ = 5.6930(9) Å, $b$ = 7.6182(16) Å, and $c$ = 5.340(2) Å at 298 K. As shown in Fig. 6(b), the central $Ti^{3+}$ ion forms a $TiO_6$ octahedron with six $O^{2-}$, and eight $Y^{3+}$ ions are located at the vertices of a distorted parallelepiped. O1 and O2 are located at the 4*c* site and 8*d* site (general position), respectively. The subscripts x, y, and z, attached to the labels of atoms in Fig. 6(b), correspond to the closest quantization axes defining the wavefunction for the Ti-3$d^1$ orbital. The site symmetry for $Ti^{3+}$ orbitals is $\bar{1}$. Therefore, even when omitting energetically higher $e_g$ ($|x^2 - y^2\rangle$ and $|3z^2 - r^2\rangle$) components, the wavefunction for the Ti-3$d^1$ orbital should be expressed as a linear combination of all three $t_{2g}$ ($|xy\rangle$,

|$yz\rangle$, and |$zx\rangle$) bases. However, most experimental studies employ a simplified model that omits one of the three components of $t_{2g}$ and that therefore satisfies a four-fold rotational symmetry (Itho *et al.*, 1999; Ichikawa *et al.*, 2000; Nakao *et al.*, 2002; Kibalin *et al.*, 2021; Voufack *et al.*, 2019). Although a recent study reported the third component (Kitou *et al.*, 2020), the obtained EDD requires improvements, as was described in the previous sections. In the present work, the third component of $t_{2g}$ is determined by XAO analysis (Tanaka *et al.*, 2008; Tanaka & Takenaka, 2012), with the minor modifications explained in Section 4.2. The refinement was carried out on $F$ by the program *REFOWF* (Sakakura, 2017) written in C++ with the *Boost C++ Library* (Boost.org).

### 3.3. Experiment A: effects of MD on EDDs

Experiment A compares the two datasets measured by the APD detector (Kishimoto & Seto, 2007) with MDA and without MDA, *i.e.* non-MDA. The MDA data were measured at four-circle angles calculated by *IUANGLE* (Tanaka *et al.*, 1994), and the non-MDA data were measured at bisecting positions. *IUANGLE* searches for the optimum four-circle angles with $\psi$-rotation using the theory originally proposed by Moon & Shull (1964) and modified by Tanaka & Saito (1975).

Other experimental settings were kept identical between the two datasets. The entire reciprocal space was measured, and eight symmetry equivalents were measured for most reflections. Other details are summarized in Table 1.

### 3.4. Experiment B: effects of statistical error on EDDs

Experiment B compares the two datasets with MDA collected by the APD detector (Kishimoto & Seto, 2007) and by a Na-I(Tl) SD. The highest count rate with negligible count loss $I_{saturate}$ is ~$10^8$ cps for the APD detector and ~$10^5$ cps for the SD. In the dataset obtained by the SD, the incident X-ray beam was attenuated by a metal foil to avoid count saturation. For large-$F$ reflections, an additional metal foil attenuating the diffracted X-rays was inserted. Other experimental settings were kept identical between the two datasets.

Since Experiment B was performed with a different beam time to Experiment A, the dataset for the APD detector with MDA was again collected in Experiment B to set up most experimental parameters as common between the two datasets. The measured reciprocal space was 1/4, and only two symmetry equivalents were measured for most reflections. Other details are summarized in Table 1.

## 4. Refinement models

The spherical ionic model (SIM) described in Section 4.1 and the orbital wavefunction model (OWM) described in Section 4.2 were applied. SIM was applied to all four datasets in Experiments A and B. OWM was applied to the dataset with MDA in Experiment A after rejecting 15 reflections as described in Section 5.1.

### 4.1. Spherical ionic model (SIM)

The atomic EDD in SIM is modeled as:

$$\rho_{core}(\mathbf{r}) + P_{valence}\, \kappa_{valence}^3\, \rho_{valence}(\kappa_{valence}\, \mathbf{r}), \qquad (8)$$

where $\rho_{core}$ is the core EDD, and $P_{valence}\, \kappa_{valence}^3\, \rho_{valence}$ is the spherical valence EDD. $P_{valence}$ is the electron population of $\rho_{valence}$, and $\kappa_{valence}$ is the expansion and contraction parameter (Coppens *et al.*, 1979) of $\rho_{valence}$. The electron configurations of $\rho_{valence}$ for atoms are chosen as $(4s)^2(4p)^6$ for $Y^{3+}$, $(3s)^2(3p)^6$ for $Ti^{3+}$, and $(2s)^2(2p)^6$ for $O^{2-}$. The spherical valence term with $\kappa_{valence}$ in equation (8) is effective for expressing the expansion and contraction of $\rho$ in a spherical ion whose radial expansion changes drastically by its valence charge and its coordination number. The Ti-$3d^1$ orbital is not included in SIM, and this can highlight the Ti-$3d^1$ orbital by difference Fourier synthesis as explained in Section 5.2. All the radial models in $\rho$ were constructed from the orbital wave functions calculated by Clementi and Roetti (1974) for neutral atoms.

Thermal motions of all atoms were treated by the anisotropic harmonic ADPs, and the extinction effect was refined by the isotropic type-I Lorentzian distribution modeled by Becker & Coppens (1974*a*, 1974*b*).

### 4.2. Orbital wavefunction model (OWM)

The omitted Ti-3$d^1$ orbital in SIM is additionally included in OWM. The following EDD model with Ti-3$d^1$ orbitals based on the formalism of XAO analysis (Tanaka *et al.*, 2008; Tanaka & Takenaka, 2012) was applied for Ti$^{3+}$:

$$\kappa_{core}^3 \rho_{core}(\kappa_{core}\,\mathbf{r}) + P_{valence}\,\kappa_{valence}^3\,\rho_{valence}(\kappa_{valence}\,\mathbf{r}) + P_{orbital}\,|\psi_{orbital}(\mathbf{r})|^2, \qquad (9a)$$

where

$$\psi_{orbital}(\mathbf{r}) = \Sigma_i\, c_i\, |i\rangle, \qquad (9b)$$

and $|i\rangle$ is a basis function. $|i\rangle$ is given by the following equation for the 3$d$-orbital cases:

$$|i\rangle = (\kappa_{3d})^{3/2} R_{3d}(\kappa_{3d} r)\, i/r^2(\theta,\varphi), \qquad (10)$$

where $i$ corresponds to $xy$, $yz$, or $zx$ for $t_{2g}$, and $x^2 - y^2$ or $3z^2 - r^2$ for $e_g$. In OWM, $\kappa_{core}$ was also introduced to the core. While equation (9a) formally has an additional spherical valence term against the original XAO formalism (Tanaka *et al.*, 2008; Tanaka & Takenaka, 2012) (see equations (26) and (27) in the original XAO paper (Tanaka *et al.*, 2008)), equation (9a) and the original formalism (Tanaka *et al.*, 2008; Tanaka & Takenaka, 2012) are mathematically identical. In the original XAO formalism, spherical valences are realizable by introduction of restrictions on parameters in the term for the wavefunction $P_{orbital}\,|\psi_{orbital}(\mathbf{r})|^2$. *REFOWF* (Sakakura, 2017) explicitly implements equation (9a). Other details are the same as for SIM.

### 5. Data analysis and evaluation method for *δF*/*F*

All four datasets from Experiments A and B were refined by SIM. OWM was applied only for the MDA dataset measured by the APD detector in Experiment A. Before application of OWM, 15 reflections were rejected using the criterion explained in Section 5.1. The results are summarized in Table 2. The accuracy of the results was examined by a combination of the difference Fourier synthesis explained in Section 5.2 and the evaluation of $\delta F/F$ described in Section 5.3.

## 5.1. Rejection criterion for outliers

Fifteen measured reflections satisfying the following equation were rejected:

$$\frac{|F_{obs} - \langle F_{obs}\rangle|}{\langle F_{obs}\rangle} > 40 \cdot \frac{0.5}{k_{sc}\langle F_{obs}\rangle} + 0.02, \tag{11}$$

where $\langle F_{obs}\rangle$ is the mean among the symmetry equivalents. The term on the left-hand side of equation (11), $|F_{obs} - \langle F_{obs}\rangle|/\langle F_{obs}\rangle$, corresponds to the observed $\delta F/F$, and the right-hand side of equation (11) expresses the significantly larger $\delta F/F$ compared to the system-dependent $\delta F/F$: 40 times larger $\delta F/F$ from Poisson statistics, *i.e.*, $40 \cdot 0.5/(k_{sc}\langle F_{obs}\rangle)$, plus 2% of baseline fluctuation. Since 40 is a considerably large number comparing to 3 (which is 99.7% coverage probability for normal distribution), and 2% is also large enough as a baseline fluctuation, equation 11 is preferable for a rejection criterion.

## 5.2. Visualization of Ti-3$d^1$ orbitals

The accuracy of the dataset can be evaluated by the following difference Fourier synthesis, which can highlight the Ti-3$d^1$ orbital:

$$\Delta\rho(\boldsymbol{r}) = \mathcal{F}^{-1}\left[\frac{F_{obs}}{F_{calc}}\boldsymbol{F}^0_{calc} - \boldsymbol{F}^0_{calc}\right], \tag{12}$$

where $\boldsymbol{F}^0_{calc}$ is $\boldsymbol{F}_{calc}$ without the anomalous dispersion terms $f'$ and $f''$, and the term $(F_{obs}/F_{calc})\boldsymbol{F}^0_{calc}$ is the approximation of $\boldsymbol{F}^0_{obs}$. Here, $\boldsymbol{F}^0_{calc}$ is modeled by SIM, excluding the Ti-3$d^1$ orbital.

### 5.3. Visualization of $\delta F/F$ for each reflection

The accuracy of individual reflections was visualized and evaluated by a simultaneous plot of $(F_{obs} - \langle F_{obs} \rangle)/\langle F_{obs} \rangle$ and $\pm 3\delta F_{estimated}/F$ against $\langle F_{obs} \rangle$. The present study defines $3\delta F_{estimated}/F$ by the following equation:

$$\frac{3\delta F_{estimated}}{F} = \frac{1.5}{k_{sc} \langle F_{obs} \rangle} + 0.01, \qquad (13)$$

where $1.5/k_{sc}$ is the approximation of $3\sigma(F_{obs})$, and 0.01 is the baseline fluctuation of $3\delta F/F$, which is roughly estimated from the distribution of $(F_{obs} - \langle F_{obs} \rangle)/\langle F_{obs} \rangle$ against $\langle F_{obs} \rangle$. (This means that $\sigma(M_{sc})/(2M_{sc})$ in the present measurement is estimated as ~0.003, and the magnitude is applied to the weight in the least-squares refinement (see Table 2).)

If the condition $|F_{obs} - \langle F_{obs} \rangle|/\langle F_{obs} \rangle > 3\delta F_{estimated}/F$ is satisfied, the reflections may contain significantly large systematic errors (including poor modeling of EDDs). As the quote from Spagna & Camalli (1999) in Section 2.3.2 indicates, no one can know true weight $(1/\delta F)^2$, since no one can know true $F$. Therefore, making goodness-of-fit approach unity is inappropriate. To make this kind of refinement rationalize, one should establish suitable model for $\delta F_{estimated}$ treating MD and other unknown errors. In the present study, we reduce $\delta F$ due to MD by avoiding MD, and the model of $\delta F_{estimated}/F$ is reduced to equation (7) with a small magnitude of $\sigma(M_{sc})/(2M_{sc})$ estimated as ~0.003. Since the remaining contribution to $\delta F$ from *Umwaganregung* process of MD and other unknown errors are not modeled in equation (7), adjusting the term $\sigma(M_{sc})/(2M_{sc})$ in equation (7) to make goodness-of-fit unity is inappropriate. For future improvements of EDD measurements, correction of the remaining causes of $\delta F$ are significant.

## 6. Results

### 6.1. Results of Experiment A: comparison of MDA and non-MDA

The results for MDA and non-MDA in Experiment A are shown in Fig. 7. Figure 7(a) shows points for $(F_{obs} - \langle F_{obs} \rangle)/\langle F_{obs} \rangle$ and lines for $\pm 3\delta F_{estimated}/F$ against $\langle F_{obs} \rangle$. The green lines are $\pm 3\delta F_{estimated}/F$ from equation (13) with $k_{sc}$ = 26.508(2), and correspond to the irreducible error level in Experiment A. The blue points are for MDA and the purple points are for non-MDA. The inset magnifies ($F_{obs}$ − $\langle F_{obs} \rangle$)/$\langle F_{obs} \rangle$ in the range −0.07 to 0.07.

Residual EDDs synthesized after refinement by SIM are shown in Figs. 7(b) for MDA and 7(c) for non-MDA. Whereas Fig. 7(b) for MDA exhibits the Ti-$3d^1$ orbital clearly, Fig. 7(c) for non-MDA failed at detection of the Ti-$3d^1$ orbital. Thus, the reduction of $\delta F/F$ by MDA is confirmed.

Figure 8 compares $F_{obs}$ and $F_{calc}$ for MDA and non-MDA. Figures 8(a) and 8(b) plot $F_{obs}$ versus $F_{calc}$ for non-MDA and MDA, respectively. Figure 8(c) compares $F_{calc}$ with and without MDA, and Fig. 8(d) compares $F_{obs}$ with and without MDA, including the systematic extinction (forbidden) reflections due to $n$-glide ($0kl$ satisfying $k + l = 2n+1$) and $a$-glide ($hk0$ satisfying $h = 2n + 1$) planes. The green lines are $\pm 3\delta F_{estimated}/F$ given by equation (13). Whereas the difference in $F_{calc}$ between MDA and non-MDA is mostly within $\pm 3\delta F_{estimated}/F$ (see Fig. 8(c)), the intensity of $F_{obs}$ increased significantly in non-MDA, owing to the contribution from the *Umweganregung* process of MD (see Fig. 8(a)). In contrast, $F_{obs}$ in MDA gained in intensity only moderately (see Fig. (b)). Figure 8(d) exhibits that most $F_{obs}$ are smaller in MDA than in non-MDA including the forbidden reflections. The result that most small-$F$ reflections are severely increased their $F_{obs}$ for non-MDA measurement including the forbidden reflections proves that this gaining is caused by *Umweganregung* process of MD whose mechanism is explained in Section 2.2.3 (for more quantitative calculations see Moon & Shull 1964 (for basic approach); Rssmanith's lifetime works, *e.g.* 2007 (for sophisticated simulation, since 1985 she wrote many papers). However, there is room for improvement in MDA, since $F_{obs}$ for a certain proportion of forbidden reflections is larger in non-MDA than in MDA, and its magnitudes significantly exceeds the region bounded by $\pm 3\delta F_{estimated}/F$. One might think that the intensity correction of *Umweganregung* process of MD proposed by Le Page & Gabe (1979) can be applicable, their correction is roughly calculating the averaged contribution of *Umweganregung* process of MD for a crystal whose $\mu r$ is zero using reflectivities averaged over reflections at a certain (sliced region of) $2\theta$. For correction of MD remaining after application of MDA, more precise simulation of MD refinable for the present experimental condition is required. As far as we know, one of the most precise and accurate simulation programs is *UMWEG* (Rossmanith, *eg*. 2007). However, for correction of *Umweganregung* process of MD, many factors should be adjusted and refined (see Conclusions of Rossmanith (2007) that *'modulus and phase of the structure factors involved, Lorentz factors, temperature factors, path lengths, polarization factors, choice of the normalized distribution function, divergence and wavelength spread of the incident beam, shape and mosaicity of the sample cosine of the angle between the electric vectors of the primary and Umweg waves, absorption , primary and secondary extinction and last but not least the approximation for Ro-06-(20) $\varphi_{lattice}(\psi)$'*). Therefore, careful implementation and development for correction of MD is needed.

This situation also accounts for inappropriateness of application of the following simple equation to remaining contribution of MD in $\delta F$ such employed in *SADABS* (Bruker, 2001) (for the following equation, see Krause *et al*., 2015): $I_{corrected} = I_{raw}S(n)P(u, v, w)Q(\mu r, 2\theta)$, where $S(n)$ is a scale factor of the $n$-th frame, $P(u, v, w)$ is a term refined to make intensities of symmetry equivalents equal by refinement of the terms expanded with spherical harmonics defined for direction cosines $u$, $v$, $w$ of the diffracted beam relative to crystal-fixed axes (Blessing, 1995), and $Q(\mu r, 2\theta)$ is a spherical absorption factor. For diagnosis of the remaining contribution of MD to $\delta F$, refining parameters to make the symmetry equivalents equal is not appropriate. Since $F_{obs}$ for small $F$-reflections mostly increased their modulus by *Umwaganregung* process of MD and satisfies $F_{obs} >$ *'the true F'* (see Figs. 8(a) and 8(b), or, *e.g*., Le Page & Gabe, 1979), $\langle F_{obs}\rangle$ calculated from the equivalents is also increased severely. Thus, finding outlier mostly rejects relatively good reflections less affected by the MD. Even after avoidance of MD, the present study did not take average over the equivalents for diagnosis of $\delta F$.

In the refinements, all the forbidden reflections were rejected.

### 6.2. Results of Experiment B: comparison of APD detector and SD

The results collected by the APD detector and an SD in Experiment B are shown in Fig. 9. Figure 9(a) shows plots of $(F_{obs} - \langle F_{obs}\rangle)/\langle F_{obs}\rangle$ and $\pm 3\delta F_{estimated}/F$ against $\langle F_{obs}\rangle$. The lines for $\pm 3\delta F_{estimated}/F$ with different $k_{sc}$ are shown in different colors: green is for $k_{sc} = 26.491(4)$ with the APD detector, blue is for $k_{sc} = 2.0561(19)$ with the SD with one attenuator, and yellow is for $k_{sc} = 0.7698(12)$ (=($k_{sc}$ *for SD with one attenuator*)/(*Attenuation factor*)$^{1/2}$) with the SD with two attenuators. The inset magnifies $(F_{obs} - \langle F_{obs}\rangle)/\langle F_{obs}\rangle$ in the same manner as Fig. 7. Residual EDDs synthesized after refinement by SIM are shown in Figs. 9(b) and 9(c) for the APD detector and the SD, respectively. Due to the large error level of Poisson statistics, the data obtained by the SD failed to detect Ti-$3d^1$ orbitals.

The reason for the failure is explained semi-quantitatively in Fig. 10. Figure 10 compares $(\boldsymbol{F}_{orb} \cdot \boldsymbol{F})/|\boldsymbol{F}|^2$ for Ti-$3d^1$ orbitals and $\pm 3\delta F_{estimated}/F$ contributed by the data collection equipment. The lines are the same as those in Fig. 9(a). Among the 1734 independent reflections, the number of reflections satisfying the condition $|\boldsymbol{F}_{orb} \cdot \boldsymbol{F}|/|\boldsymbol{F}|^2 > 3\delta F_{estimated}/F$ is only 5 for the blue lines for the SD with one attenuator. The corresponding number for the green lines for the APD detector is 116, which amounts to 6.7% of all the independent reflections.

### 6.3. Observed Ti-$3d^1$ orbitals

The orbital wavefunction refined for the dataset of MDA in Experiment A with rejection of 15 reflections by equation (11) is $0.130(3) |xy\rangle + 0.705(2) |yz\rangle + 0.698(2) |zx\rangle$. Figure 11 shows the EDDs for the refined Ti-$3d^1$ wavefunction. Reflecting the site symmetry $\bar{1}$, two of the four $3d$ lobes are more elongated towards the closest $Y^{3+}$. This suggests considerable stabilization due to electrostatic attraction from the closest $Y^{3+}$. Although in the present results, the degree of elongation of the lobes is reversed compared to a theoretical calculation (Pavarini, 2004) between the 1st and 2nd elongated lobes, discussion on this point is beyond the scope of the present paper and will appear in our future paper.

### 7. Discussion

Since the results section already confirms most topics on $\delta F$ introduced in the theoretical section, the present section is restricted to the topic of residual EDDs from the refinement by OWM.

Figures 12(a) and 12(b) describe $\Delta\rho$ in equation (12) before and after refinement by OWM, respectively. Although refinement by OWM erases the highest peak of 2.06 eÅ$^{-3}$ corresponding to the lobes of the Ti-$3d^1$ orbital, the highest peak of 0.99 eÅ$^{-3}$ still remains near the Ti atom. The EDDs could not be eliminated even by another refinement model adding two more bases of $e_g$ (the results will appear in our future paper). The systematically concentrated residual EDDs around Ti with 0.99 eÅ$^{-3}$ strongly suggest that the residual EDDs are not an artifact but are due to $3d$-electrons.

Thus, the following two causes are plausible. One is the limitation of the OWM with a linear combination of atomic orbitals (LCAO) where bases are expressed by variable separation, since the shape of the surrounding potential cannot be expressed separately along the radial and angular directions. The other is the partial occupation of another energetically higher $3d$ orbital. In either case, construction of an appropriate EDD model is not easy and needs further investigation. However, with further improvements in data quality, we are certain to be able to reveal the true behavior of electrons.

Figure 12(c) shows $(F_{obs} - F_{calc})/F_{calc}$ after refinement by OWM against $F_{calc}$. Since a large proportion of $(F_{obs} - F_{calc})/F_{calc}$ exceeds the line $+3\delta F_{estimated}/F$ at small $F_{calc}$, a systematic gain in $F$ due to the *Umweganregung* process of MD remains. If these reflections are rejected or measured with more accuracy, we can obtain more accurate EDDs. This topic will also be addressed in a future paper.

## 8. Summary and Prospects

The present paper discussed the following topics. (A) All aspherical wavefunctions other than *s* have nodal planes passing through the nuclear position. Therefore, higher-angle reflections generally contain information on anisotropy near the nucleus, which becomes clearer when the smearing due to thermal motion of atoms is reduced. On the other hand, lower-order reflections contain information on orbital anisotropies far from the nucleus, and the smearing due to thermal motion is generally moderate. (B) $\delta F/F$ in real experiments was discussed and the necessity of separate treatments of statistical and systematic errors was presented. In most statistical optimization methods including least-squares refinements, $\delta F/F$ from statistical causes can be treated in the weight for each reflection. However, $\delta F/F$ from systematic causes should be treated using a model that refines the data as a correction factor. Thus, the effects of MD and counting statistics are better treated separately, and weighting schemes to make the goodness-of-fit approach unity are an unsuitable approach in some cases. (C) By $\delta F/F$ reduction, the present study reduced $\delta \rho$, and orbital anisotropies greater than ~0.4 Å from the nuclear position were clearly observed. However, the results shown in Fig. 12(b) indicate the difficulties in constructing mathematical models for orbitals in the vicinity of the Fermi level, and $(F_{obs} - F_{calc})/F_{calc}$ after refinement by OWM plotted against $F_{calc}$ in Fig. 12(c) shows a remaining systematic error due to MD. Since the exceeding magnitudes of $(F_{obs} - F_{calc})/F_{calc}$ from $+3\delta F_{estimated}/F$ in Fig. 12(c) are mostly comparable to $1.5/k_{sc}$, further enlargement of $k^0_{sc}$, in addition to reduction of MD, is also required for further improvement of the observable $\rho$.

Although the present study is targeting the Ti-$3d^1$ orbital, one can also apply $|\boldsymbol{F}_{orb} \cdot \boldsymbol{F}|/|\boldsymbol{F}|^2 > 3\delta F_{estimated}/F$ for diagnosis of the observability for multi-electron orbitals. However, one should note that as the number of electrons contributing to $\boldsymbol{F}_{orb}$ increases, the numerator $|\boldsymbol{F}_{orb} \cdot \boldsymbol{F}|$ approaches the denominator $|\boldsymbol{F}|^2$. Therefore, $|(\boldsymbol{F}_{aspherical\_moel} - \boldsymbol{F}_{spherical\_model}) \cdot \boldsymbol{F}|/|\boldsymbol{F}|^2 > 3\delta F_{estimated}/F$ can be another candidate for this kind of evaluation.

In strongly correlated materials, quantitative evaluations of the electron populations and the spatial distribution of valence electrons have great importance. Metallic $d$ and $f$ orbitals in most oxides often exchange electrons with the surrounding $O^{2-}$ and exhibit interactions such as super-exchange. Whereas the Goodenough-Kanamori rules (Goodenough, 1955, 1958; Kanamori, 1959) empirically explain the ferro- and antiferromagnetic arrangement of spins, capturing the anisotropy by electron density observation is still challenging. The small electron populations exchanged between the metallic cation and the ligand $O^{2-}$ are dilute EDDs, and the wide distribution of $2p$ orbitals on $O^{2-}$ (*e.g.*, 1.4 Å for the Shannon ionic radius (1976) of six-coordinated $O^{2-}$) dilute the EDD further. Therefore, further reduction of $\delta F/F$ is required, especially in the relatively small $\sin\theta/\lambda$ range (large $d$).

The present SSXRD has the potential to realize very small $\delta F/F$. Thus, what needs to be achieved is a small $\delta F/F$ and a large number of reflections satisfying $F_{\text{orb}}/F > N \cdot \delta F/F$. To this end, the current theoretical and technical standards and foundations must be revised and improved.

**Acknowledgements**

The authors would like to thank Dr. R. Kiyanagi of J-PARC, as well as Dr. S. Mitsuda, professor of physics at Tokyo University of Science, and his laboratory team for useful discussions on the topics in Section 2.2.1.

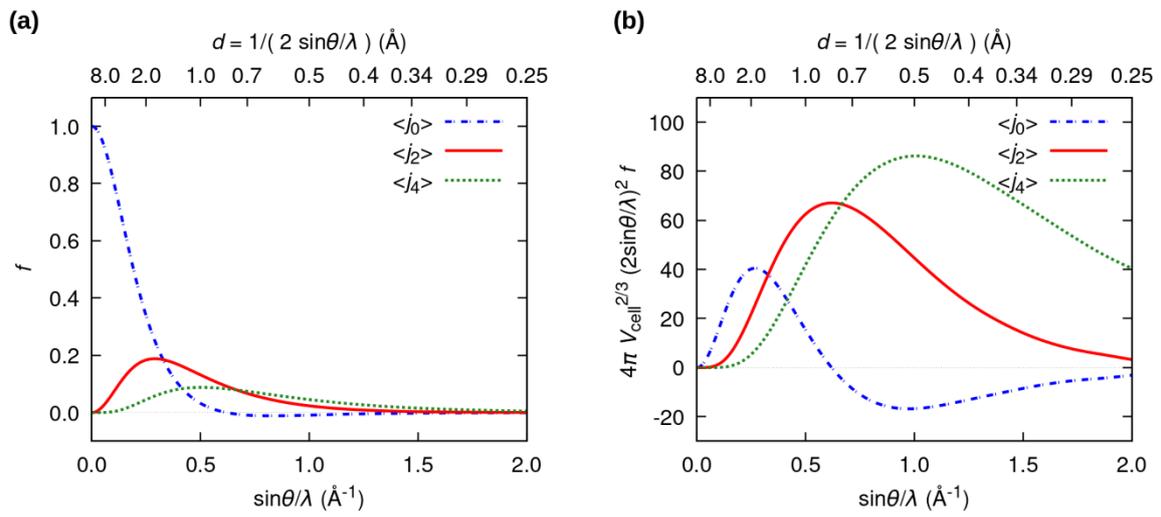

**Figure 1** Radial scattering factors for Ti-$3d^1$ electrons versus $\sin\theta/\lambda$ and $d$ $(=1/(2\sin\theta/\lambda))$. The vertical axes are $\langle j_k \rangle$ for (a), and $4\pi V_{cell}^{2/3} (2\sin\theta/\lambda)^2 \langle j_k \rangle$ for (b). For details see the text.

**Figure 2** EDDs for Ti-3$d^1$ orbitals synthesized by inverse Fourier transformation from reflections in different sin$\theta/\lambda$ ranges. The top panels, (a) to (c), do not convolute the Debye-Waller factor, and the bottom panels, (d) to (f), convolute the Debye-Waller factor, which is determined in the present study at 298 K. The ranges of sin$\theta/\lambda$ are 0 < sin$\theta/\lambda$ < 2.0 Å$^{-1}$ for (a) and (d), 0 < sin$\theta/\lambda$ < 1.2 Å$^{-1}$ for (b) and (e), and 1.2 < sin$\theta/\lambda$ < 2.0 Å$^{-1}$ for (c) and (f). All the contour intervals are 0.1eÅ$^{-3}$. The contours of the solid red lines, the broken green lines, and the dotted dark orange lines are positive, negative, and zero levels, respectively. Respective ($\rho_{max}$, $\rho_{min}$) for (a) to (f) are (1.98 eÅ$^{-3}$, -0.06 eÅ$^{-3}$), (1.57 eÅ$^{-3}$, -0.11 eÅ$^{-3}$), (0.76 eÅ$^{-3}$, -0.46 eÅ$^{-3}$), (1.54 eÅ$^{-3}$, 0.00 eÅ$^{-3}$), (1.36 eÅ$^{-3}$, -0.04 eÅ$^{-3}$), and (0.42 eÅ$^{-3}$, -0.24 eÅ$^{-3}$).

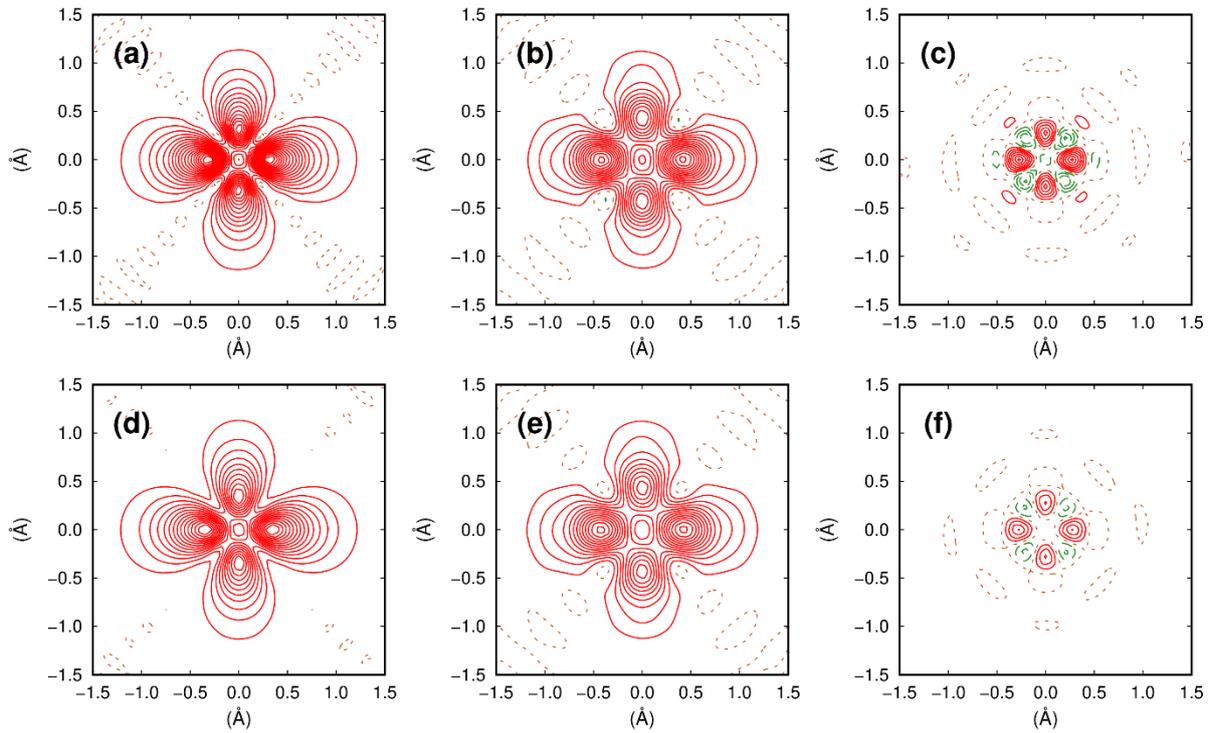

**Figure 3** $(F_{orb} \cdot F)/|F|^2$ (*i.e.*, contribution rate of Ti-$3d^1$ to the total *F*) versus $\sin\theta/\lambda$. Anomalous dispersion terms are excluded in (a) and included in (b).

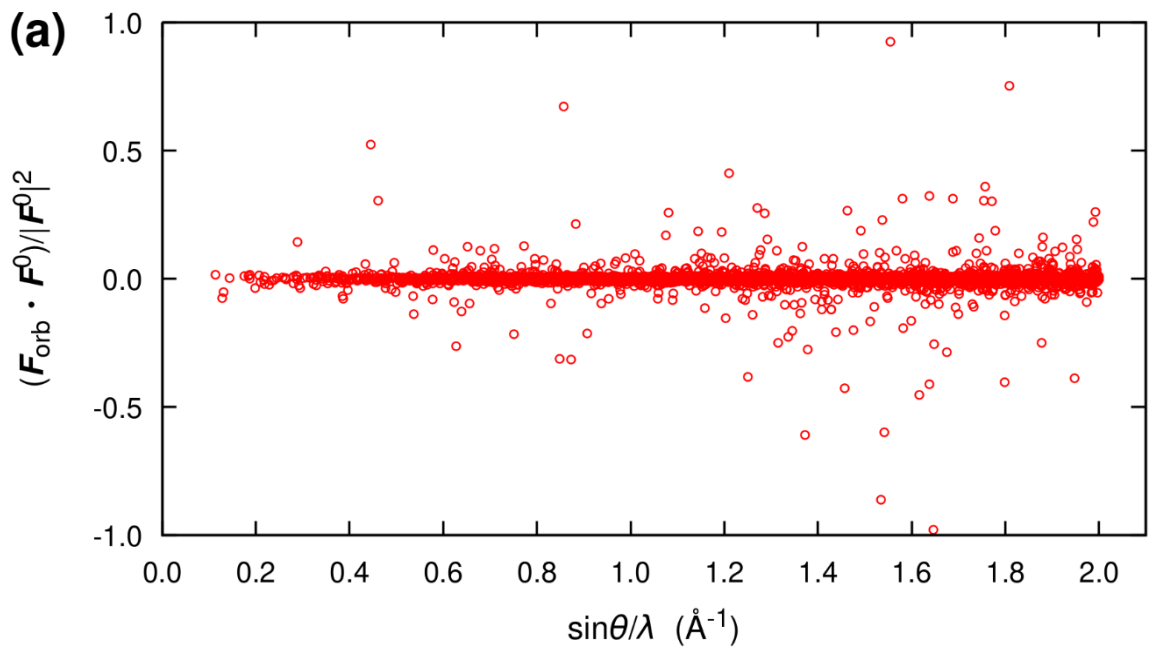
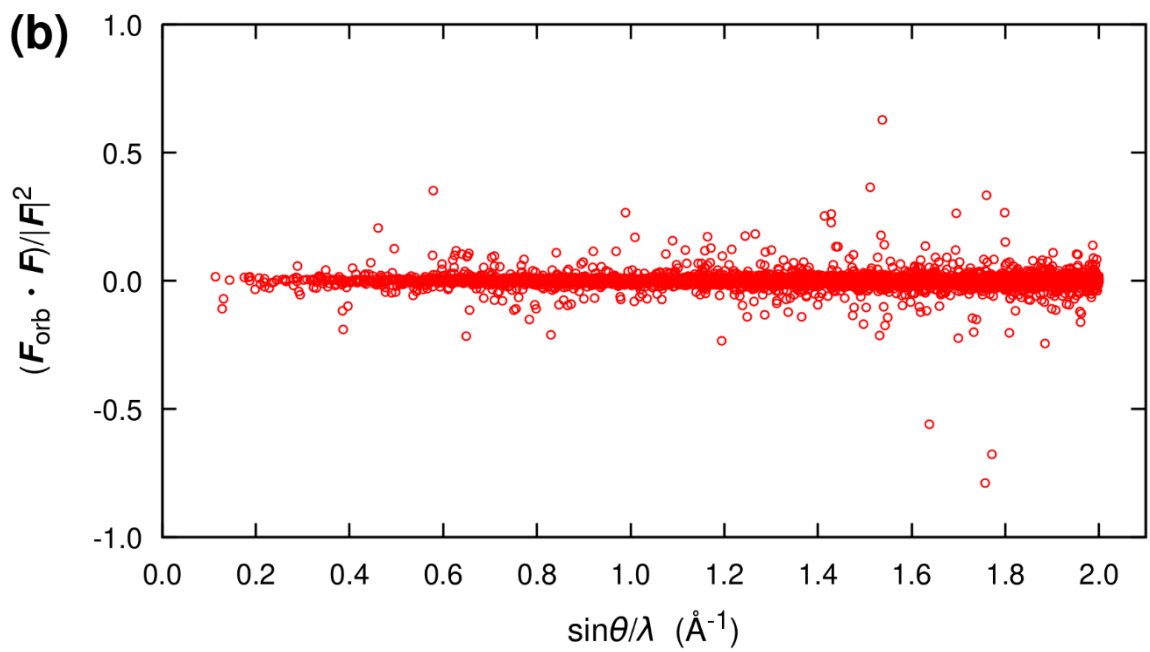

**Figure 4** (a) Graphical representation of distances in $F^1$-space and $F^n$-space. The red line is $y = (f^0)^n$, and the green line is the tangent at $(F, F^n)$. The line segments $\Delta F$ ($\equiv |F - f^0(\mathbf{p})|$), $\Delta F^n$ ($\equiv |F^n - (f^0(\mathbf{p}))^n|$) and $\Delta F_{\text{approximated}}$ ($\equiv |(F^n - (f^0(\mathbf{p}))^n)/(\partial(F^n)/\partial F)|$) are shown by blue, magenta, and brown lines, respectively. (b) The transformed Poisson distribution in $F^1$-space and its approximation by the normal distribution with $\sigma(F) = 0.5$. The expectation value is $F_0 = 2$ in both distribution functions. The Poisson distribution is shown by the red points, and the normal distribution by the broken green line. (c) The identical Poisson distribution to (b) in $F^2$-space of $I_0 = 4$ (*i.e.*, $F_0 = 2$) and its approximation by a normal distribution with $\sigma(I) = 2$. The Poisson distribution is shown by the blue points and the normal distribution by the broken magenta line.

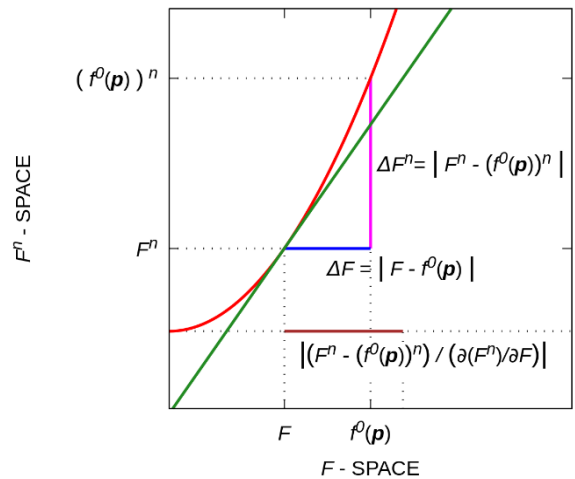

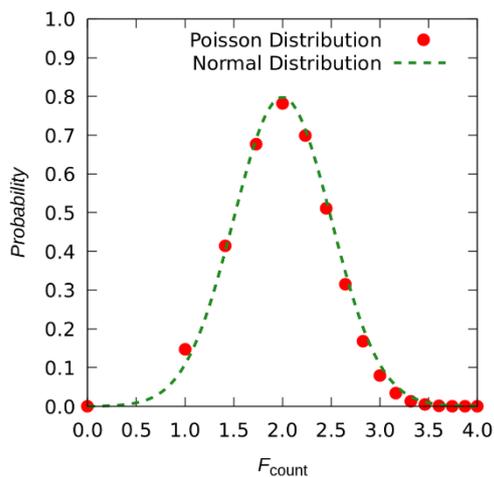
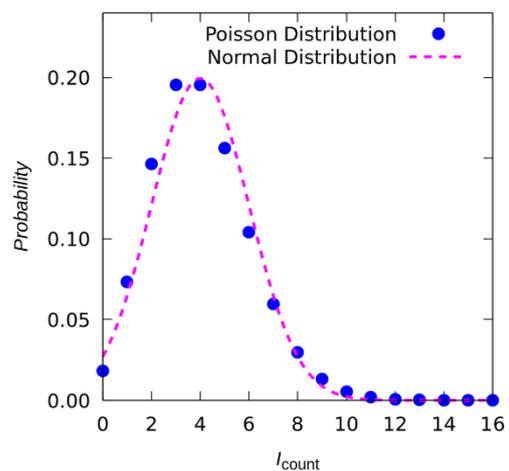

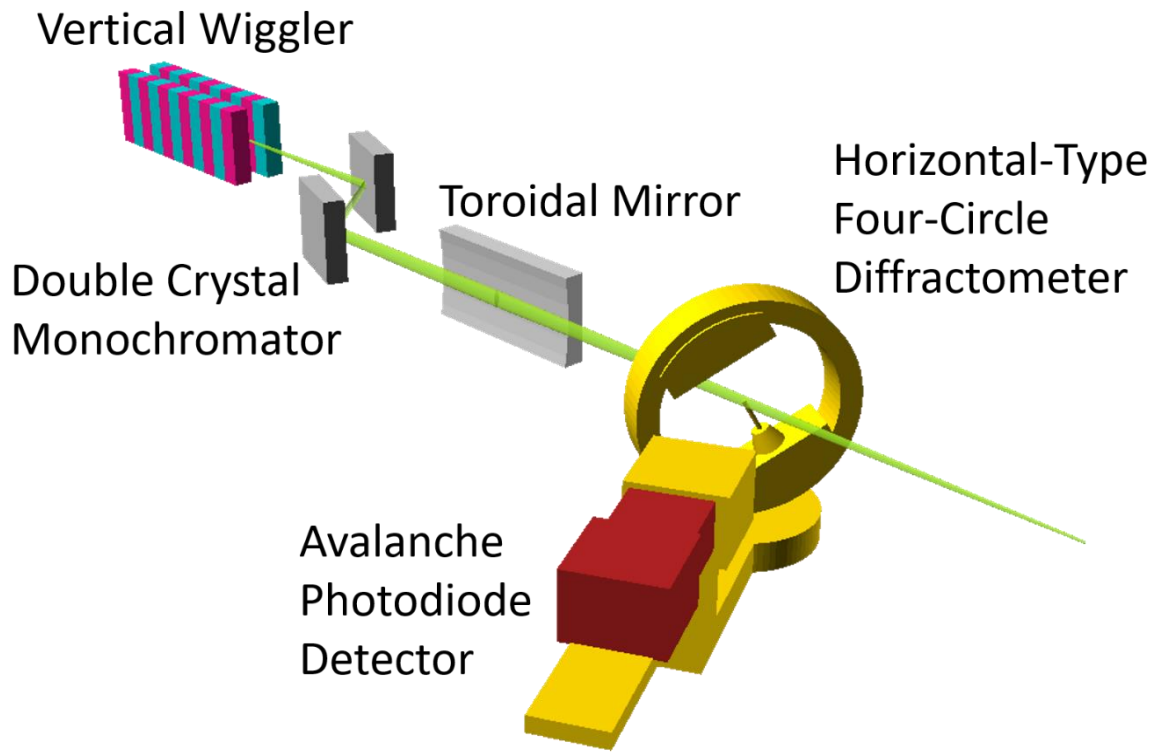

**Figure 5** Schematic of experimental station: beamline BL14A at KEK-PF (Tsukuba, Japan).

**Figure 6** Crystal structure of YTiO$_3$. (a) A unit cell structure of *Pnma* space group, and (b) a magnified view with quantization axes scaled to 3 Å. In (b), the subscripts (x, y, and z) on the labels of oxygen atoms indicate the closest quantization axes, and the subscripts (1st to 4th) on yttrium atoms indicate the closeness to the central Ti. The site symmetry of Ti is $\bar{1}$.

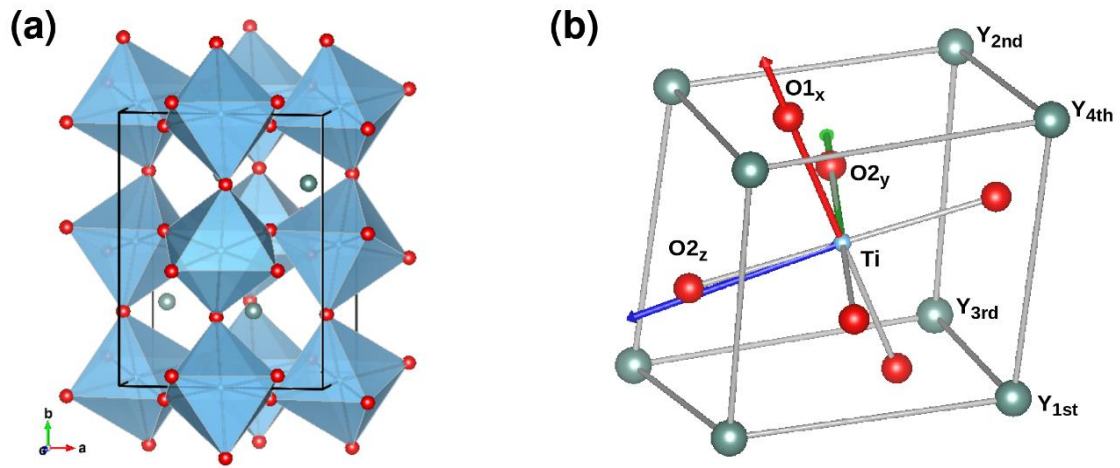

**Figure 7** Comparison of deviation of $F_{obs}$ between the data with and without MDA in Experiment A: (a) $(F_{obs} − \langle F_{obs}\rangle)/\langle F_{obs}\rangle$ versus $\langle F_{obs}\rangle$, (b) the residual EDDs for MDA, and (c) the residual EDDs for non-MDA after SIM refinement (see text). In (a), the green lines are $\pm 3\delta F_{estimated}/F$ for MDA, given by equation (13). In (b) and (c), iso-density surfaces in yellow and sky blue are $+1.1$ eÅ$^{-3}$ and $−1.1$ eÅ$^{-3}$, respectively. $(\Delta\rho_{max}, \Delta\rho_{min})$ are $(+2.08$ eÅ$^{-3}, −1.51$ eÅ$^{-3})$ for (b) and $(+3.23$ eÅ$^{-3}, −5.42$ eÅ$^{-3})$ for (c).

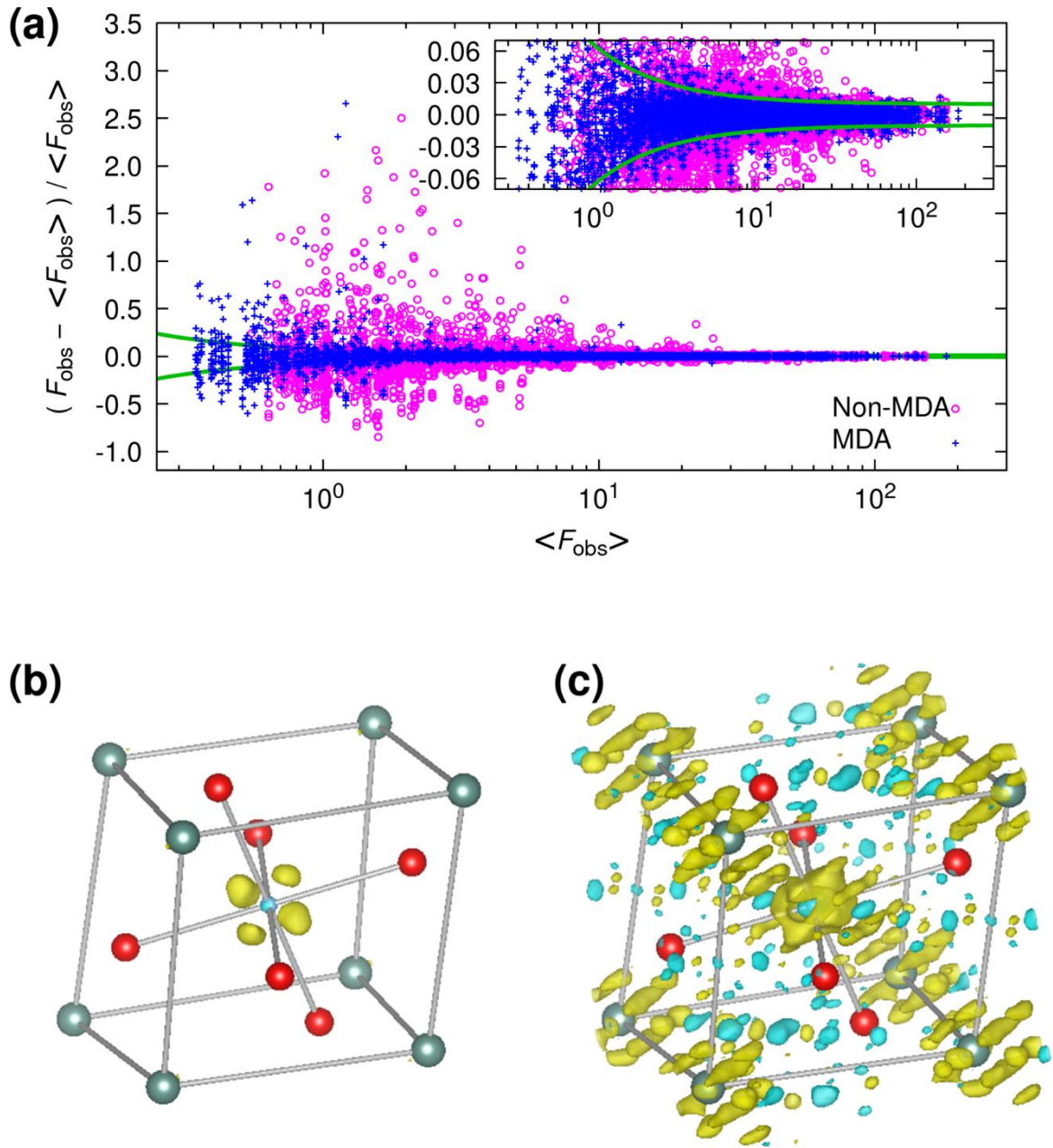

**Figure 8** Comparison of MDA and non-MDA: (a) $F_{obs}$ versus $F_{calc}$ for non-MDA, (b) $F_{obs}$ versus $F_{calc}$ for MDA, (c) $F_{calc}$ for non-MDA versus $F_{calc}$ for MDA, (d) $F_{obs}$ for non-MDA versus $F_{obs}$ for MDA including systematic extinction reflections due to $a$-glide and $n$-glide planes. The same irreducible error level shown in Fig. 7(a) is reproduced as broken green lines.

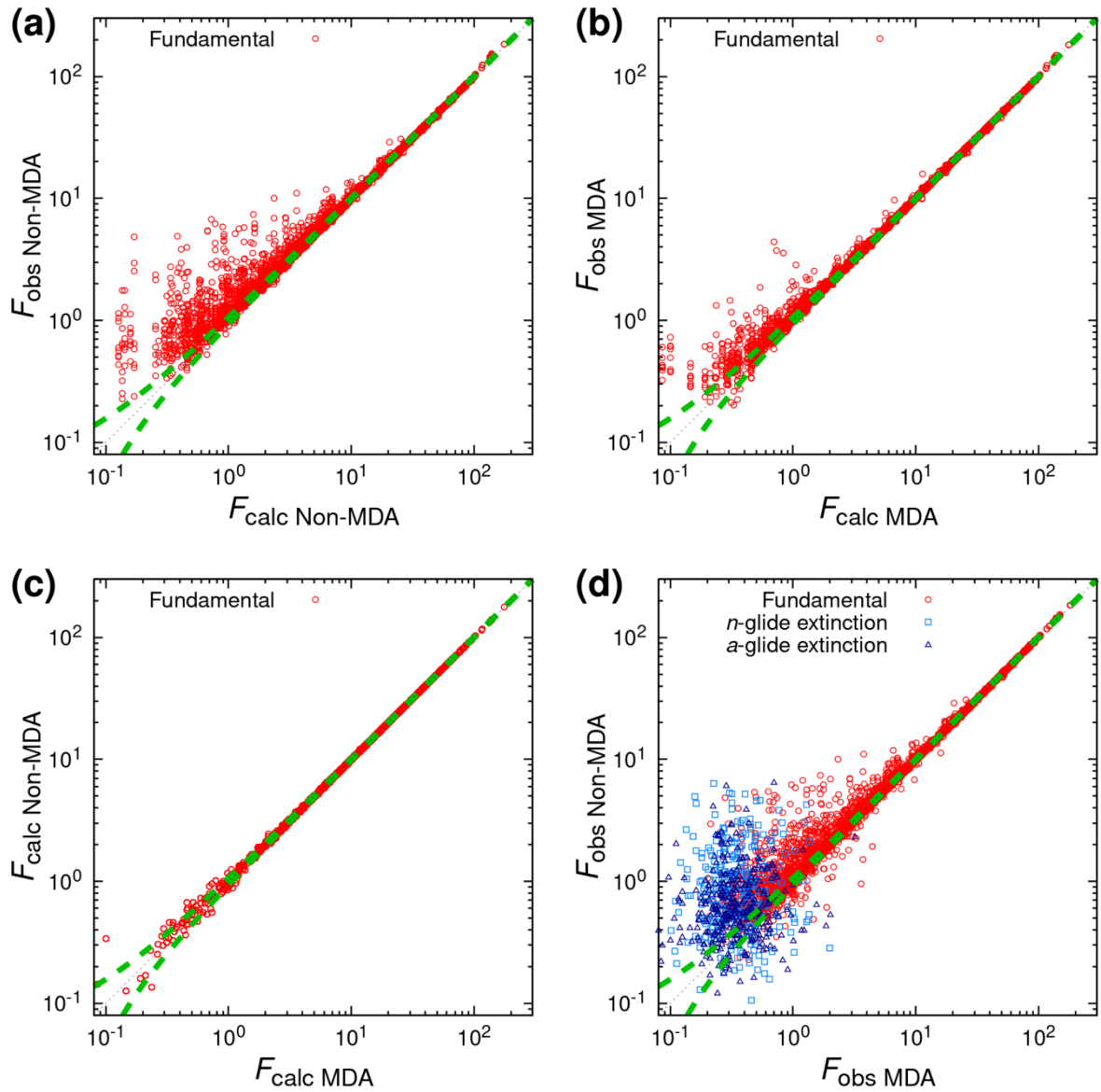

**Figure 9** Comparison of deviation of $F_{obs}$ between the data collected by the APD detector and an SD in Experiment B: (a) $(F_{obs} - \langle F_{obs} \rangle)/\langle F_{obs} \rangle$ versus $\langle F_{obs} \rangle$, (b) residual EDDs after SIM refinement (see text) for the APD detector, and (c) residual EDDs after SIM refinement for the SD. In (a), the green, blue, and yellow lines correspond, respectively, to $\pm 3\delta F_{estimated}/F$ for the APD detector, an SD with one attenuator, and an SD with two attenuators. In (b) and (c), iso-density surfaces in yellow and sky blue are $+1.1$ eÅ$^{-3}$ and $-1.1$ eÅ$^{-3}$, respectively. ($\Delta\rho_{max}$, $\Delta\rho_{min}$) are ($+1.76$ eÅ$^{-3}$, $-1.33$ eÅ$^{-3}$) for (b), and ($+2.40$ eÅ$^{-3}$, $-2.44$ eÅ$^{-3}$) for (c).

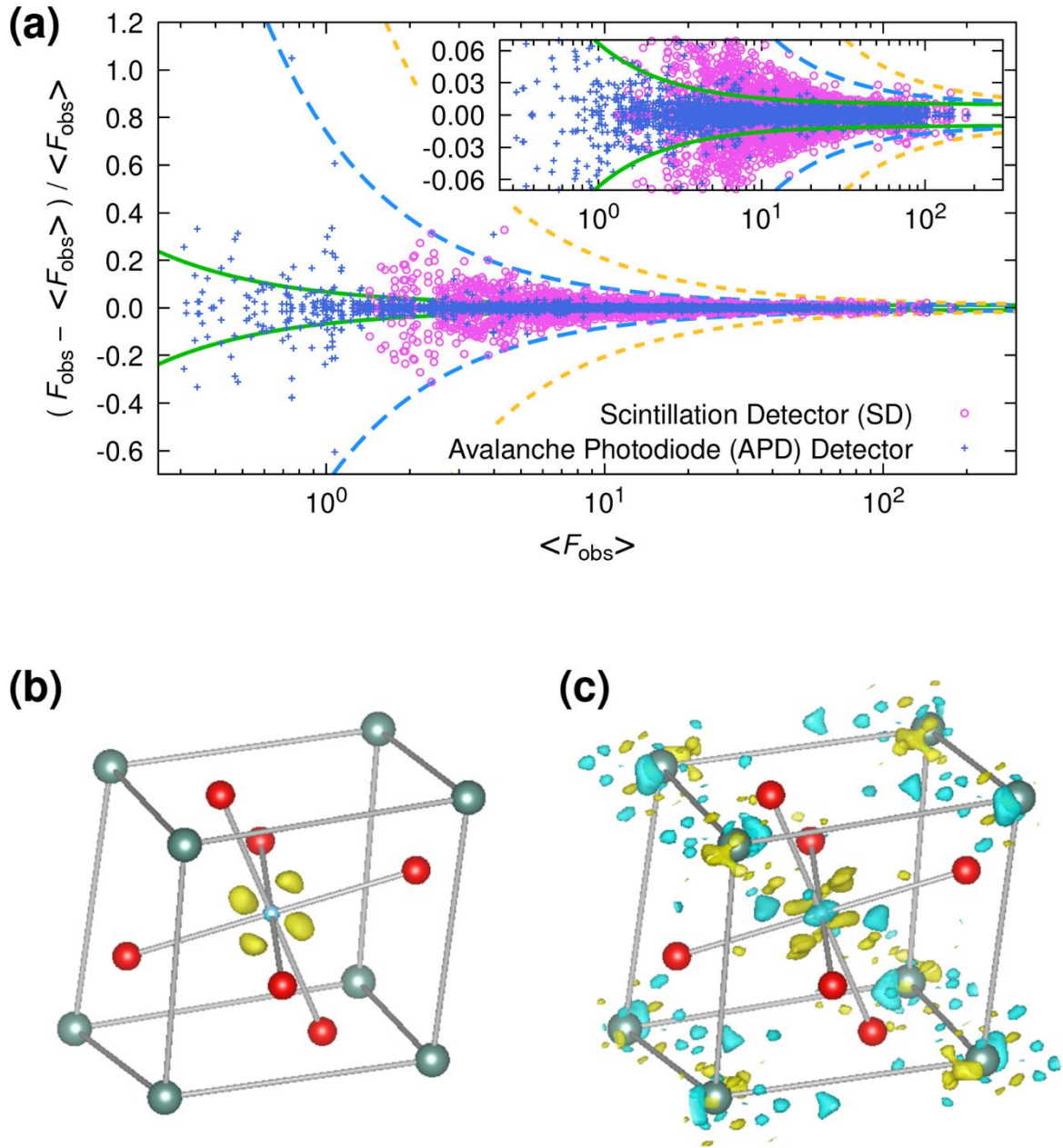

**Figure 10** $(\boldsymbol{F}_{\mathrm{orb}} \cdot \boldsymbol{F})/|\boldsymbol{F}|^2$ versus $|\boldsymbol{F}|$. The same irreducible error levels shown in Fig. 9(a) are reproduced.

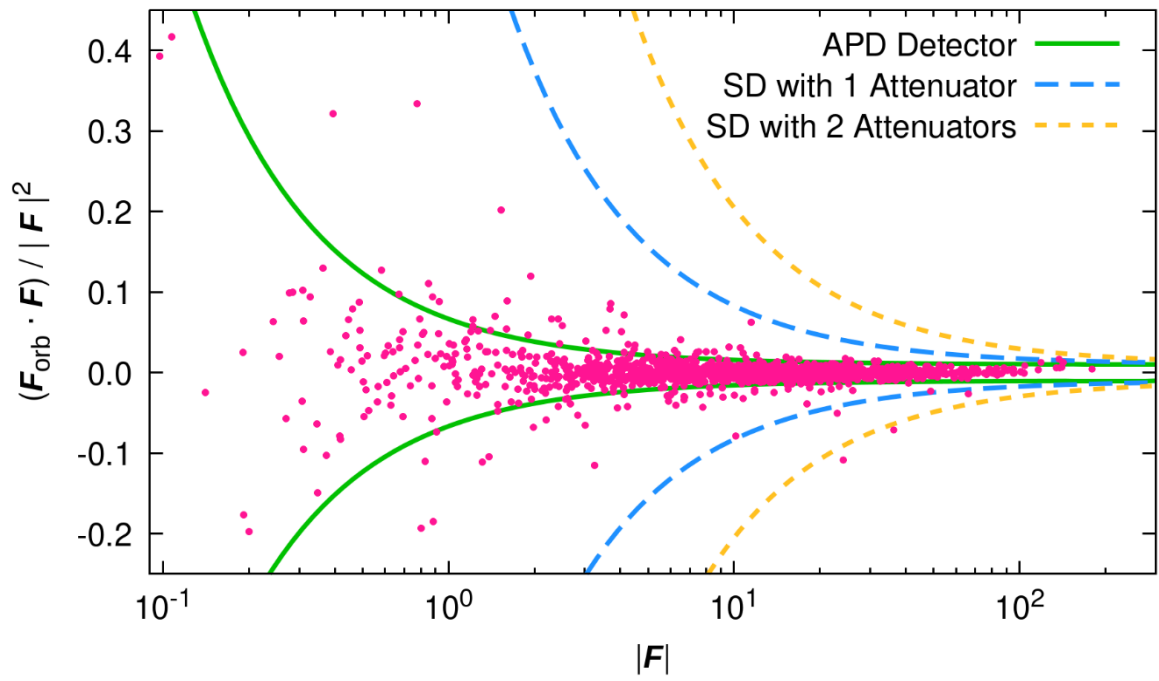

**Figure 11** EDDs for the refined orbital wavefunction of the Ti-$3d^1$ electrons in YTiO$_3$. Iso-density surfaces are +0.1 eÅ$^{-3}$ in white, +0.5 eÅ$^{-3}$ in yellow, and +1.0 eÅ$^{-3}$ in red. Arrows and the labels of Y atoms are the same as in Fig. 6(b). The sky-blue plane passes through the four density maxima of the Ti-$3d^1$ electron orbital, and the dashed green lines are the eye guide connecting the cross points of the sky-blue planes and the Y-Y segments constituting the distorted parallelepiped.

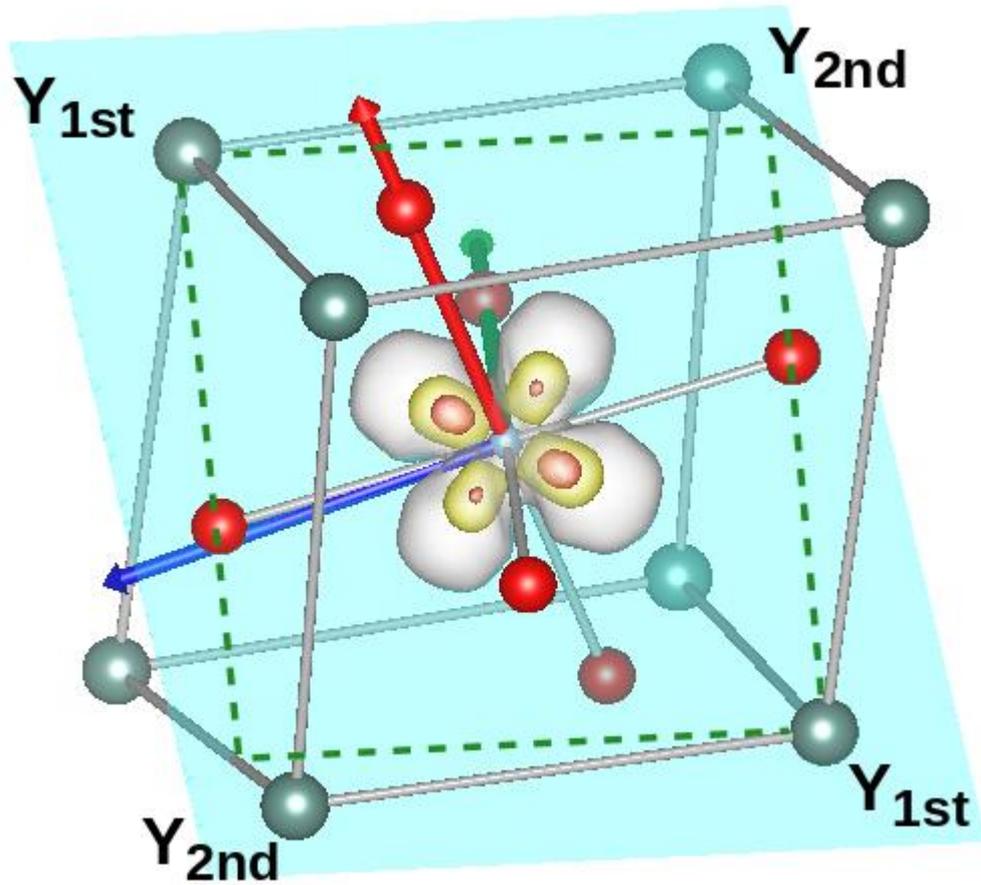

**Figure 12** Residual EDDs before and after refinement of Ti-$3d^1$ by OWM ((a) and (b)) and ($F_{obs}$ − $F_{calc}$)/$F_{calc}$ versus $F_{calc}$ after refinement (c): (a) residual EDDs before refinement, (b) residual EDDs after refinement, and (c) ($F_{obs}$ − $F_{calc}$)/$F_{calc}$ versus $F_{calc}$. ($\Delta\rho_{max}$, $\Delta\rho_{min}$) are (+2.06 eÅ$^{-3}$, −1.46 eÅ$^{-3}$) in (a) and (0.99 eÅ$^{-3}$, −0.84 eÅ$^{-3}$) in (b). The iso-density surfaces in red, yellow, and blue are, respectively, +0.90 eÅ$^{-3}$, +0.70 eÅ$^{-3}$, and −0.7 eÅ$^{-3}$. In (c), the same irreducible error level shown in Fig. 7(a) is reproduced as broken green lines.

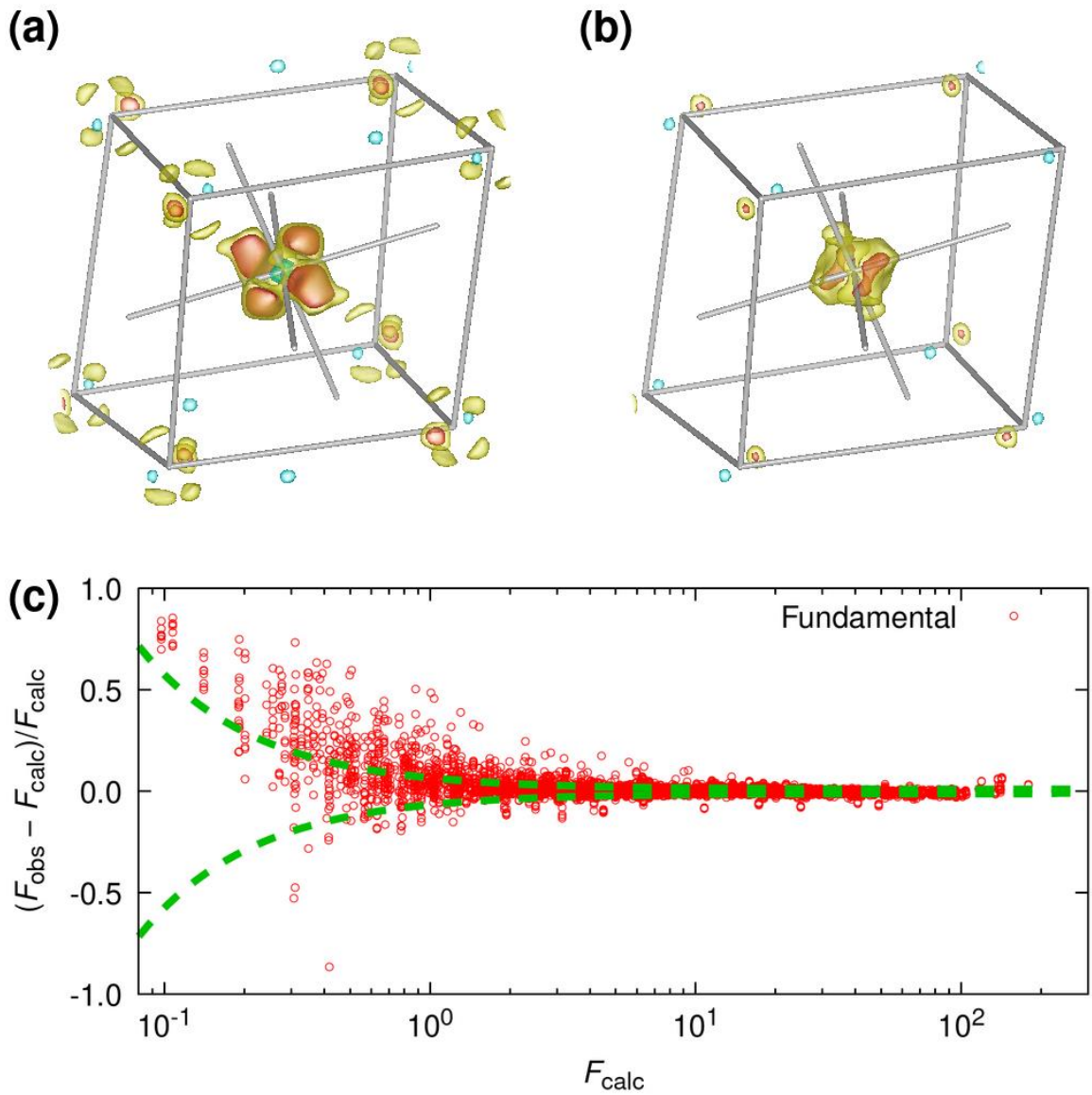

**Table 1** Experimental details of experiment A (MDA and non-MDA) and B (APD and SC).

| Common parameters | MDA & non-MDA (Experiment A) | APD & SC (Experiment B) |
|---|---|---|
| Crystal data | | |
| Chemical formula | YTiO$_3$ | |
| Crystal system, space group | Orthorhombic, *Pnma* | |
| $a$, $b$, $c$ (Å) | 5.6930(9), 7.6182(16), 5.340(2) | 5.6924(8), 7.6181(12), 5.340(2) |
| $\mu$ (mm$^{-1}$) | 8.23 | |
| Crystal shape, radius (mm) | Sphere, 0.066 | |
| Data collection | | |
| Temperature (K) | 298 | |
| Radiation type | Synchrotron, $\lambda$=0.75500(2)Å | Synchrotron, $\lambda$=0.75435(2)Å |
| Diffractometer | Four-circle@KEK-PF-BL14A (Tsukuba, Japan) | |
| Scan method | Integrated intensities from $\omega/2\theta$ scans | |
| Absorption correction | For a sphere | |
| Weight | $w = 1/(\sigma^2(F_{obs}) + 0.00009\ F^2_{obs})$ | |

**Non-common parameters in MDA, non-MDA (Experiment A), APD, and SC (Experiment B).**

| | MDA | Non-MDA | APD | SC |
|---|---|---|---|---|
| Number of measured reflections (all/independent) | 12,259 / 1,734 | 12,259 / 1,734 | 3,565 / 1,524 | 3,565 / 1,524 |
| Number of reflections used ($F>3\sigma$) (all/independent) | 12,258 / 1,734 | 12,259 / 1,734 | 3,564 / 1,524 | 3,403 / 1,477 |
| Resolution $(\sin\theta/\lambda)_{max}$ (Å$^{-1}$) | 1.200 | 1.200 | 1.150 | 1.150 |
| $T_{max}$, $T_{min}$ | 0.4536, 0.4931 | 0.4536, 0.4931 | 0.4536, 0.4894 | 0.4536, 0.4894 |
| $R_{int}(F)$ ($\equiv \Sigma|F_{obs} - \langle F_{obs}\rangle|/\Sigma\langle F_{obs}\rangle$) (%) | 0.448 | 1.004 | 0.336 | 0.990 |
| $R_{int}(F^2)$ ($\equiv \Sigma|F^2_{obs} - \langle F^2_{obs}\rangle|/\Sigma\langle F^2_{obs}\rangle$) (%) | 0.778 | 1.176 | 0.630 | 1.443 |
| $R_{int}(I)$ ($\equiv \Sigma|I_{obs} - \langle I_{obs}\rangle|/\Sigma\langle I_{obs}\rangle$) (%) | 0.815 | 1.163 | 0.679 | 1.464 |
| Goodness of fit ($\equiv (\{\Sigma w(F_{obs} - F_{calc})^2\}/(N_{refl} - N_{param}))^{1/2}$) | 4.76 | 12.77 | 4.98 | 2.00 |
| $k_{sc}$ | 26.508(2) | 26.641(2) | 26.491(4) | 2.0561(9) |
| Attenuation factor | - | - | - | 7.134(11) |
| $R(F)$ ($\equiv \Sigma|F_{obs} - F_{calc}|/\Sigma F_{obs}$) (%) | 0.975 | 1.726 | 1.045 | 1.869 |
| Number of parameters | 33 | 33 | 33 | 33 |

**Table 2** Summary of XAO analysis.

**Summary of statistics**

| | |
|---|---|
| Number of reflections used (all/independent) | 12,243 / 1,734 |
| $R_{int}(F)$ ($\equiv \Sigma|F_{obs} - \langle F_{obs}\rangle|/\Sigma\langle F_{obs}\rangle$) (%) | 0.424 |
| $R_{int}(F^2)$ ($\equiv \Sigma|F^2_{obs} - \langle F^2_{obs}\rangle|/\Sigma\langle F^2_{obs}\rangle$) (%) | 0.767 |
| $R_{int}(I)$ ($\equiv \Sigma|I_{obs} - \langle I_{obs}\rangle|/\Sigma\langle I_{obs}\rangle$) (%) | 0.806 |
| $R(F)$ ($\equiv \Sigma|F_{obs} - F_{calc}|/\Sigma F_{obs}$) (%) | 0.792 |
| Goodness of fit ($\equiv (\{\Sigma w(F_{obs} - F_{calc})^2\}/(N_{refl} - N_{param}))^{1/2}$) | 3.49 |
| Weight | $w = 1/(\sigma^2(F_{obs}) + 0.000009\, F^2_{obs})$ |
| Number of parameters | 37 |

**Refined parameter values**

| | |
|---|---|
| $k_{sc}$ | 26.619(2) |
| $g_{iso}$ | 3589(6) |
| Atomic position parameters (x,y,z) | |
| Y1 | 0.0733125(7), 1/4, 0.0210329(9) |
| Ti1 | 0, 0, 1/2 |
| O1 | -0.042749(4), 1/4, -0.379151(5) |
| O2 | -0.190494(3), 0.058315(2), 0.190257(4) |
| Atomic displacement parameters ($U_{11}, U_{22}, U_{33}, U_{12}, U_{13}, U_{23}$) (Å$^2$) | |
| Y1 | 0.0045578(18), 0.0064664(19), 0.005774(2), 0, 0.0003713(7) |
| Ti1 | 0.003112(7), 0.003225(7), 0.003443(8), 0.0002363(11), -0.0000001(10), 0.0000139(10) |
| O1 | 0.006980(6), 0.005072(6), 0.006934(8), 0, -0.001136(6), 0 |
| O2 | 0.006229(4), 0.008228(5), 0.006407(5), -0.001496(3), -0.001389(4), 0.001039(3) |
| $\kappa$ values for the spherical ionic model ($\kappa_{core}, \kappa_{valence}$) | |
| Y1 | 1, 1.299(5) |
| Ti1 | 0.9438(5), 1.1045(4) |
| O1 | 1, 0.9134(2) |
| O2 | 1, 0.90624(14) |
| Ti-3$d^1$ valence orbital parameters | |
| Radial expansion parameter ($\kappa_{orbital}$) | 0.898(2) |
| Angular orbital wavefunction | 0.130(3)$|xy\rangle$ + 0.705(2)$|yz\rangle$ + 0.698(2)$|zx\rangle$ |
| Quantization axes defined on Ti$^{3+}$ at (0,0,1/2) ($q_x$, $q_y$, $q_z$) | (-0.11997, 0.94027, 0.31859), (-0.53375, 0.20949, -0.81928), (-0.83709, -0.26833, 0.47674) |

## Appendix A. Orbital scattering factors $\langle j_k \rangle$

The scattering factor for one-center orbitals whose angular part is a spherical harmonic function is reviewed. The following equations form the basis of the XAO analysis (Tanaka *et al.*, 2008; Tanaka & Takenaka, 2012).

Let $j_k$, $Q$, **r**, **h**, $C^k(lm,l'm')$, and $\langle j_k(Q) \rangle$ be, respectively, the spherical Bessel function, $4\pi \sin\theta_B/\lambda$, $(r, \theta, \phi)$, $(2\sin\theta_B/\lambda, \theta', \phi')$, $\langle Y_{l,m} | \sqrt{4\pi/(2k+1)} Y_{k,S} | Y_{l',m'} \rangle$, and $\langle \kappa_{nl}^{3/2} R_{nl} | j_k | \kappa_{n'l'}^{3/2} R_{n'l'} \rangle$. Then, the orbital scattering factor $f_{i,j}(\mathbf{h})$ defined by the two orbitals $|i\rangle$ ($\equiv |nlm\rangle$) and $|j\rangle$ ($\equiv |n'l'm'\rangle$) is given by:

$$f_{i,j}(\mathbf{h}) = \langle nlm | e^{2\pi i \mathbf{h}\mathbf{r}} | n'l'm' \rangle \quad (A1)$$

$$= \langle nlm | \sum_{k=0}^{\infty} i^k 4\pi j_k(Qr) \sum_{s=-k}^{k} Y_{k,s}(\theta,\phi) Y_{k,s}^*(\theta',\phi') | n'l'm' \rangle \quad (A2)$$

$$= \sum_{k=0}^{\infty} i^k \sqrt{4\pi(2k+1)} \left\langle \kappa_{nl}^{\frac{3}{2}} R_{nl}(\kappa_{nl} r) \middle| j_k(Qr) \middle| \kappa_{n'l'}^{\frac{3}{2}} R_{n'l'}(\kappa_{n'l'} r) \right\rangle$$

$$\sum_{s=-k}^{k} Y_{k,s}^*(\theta',\phi') C^k(lm,l'm') \quad (A3)$$

$$= \sum_{k=|l-l'|}^{l+l'} i^k \sqrt{4\pi(2k+1)} \langle j_k(Q) \rangle Y_{k,m-m'}^*(\theta',\phi') C^k(lm,l'm'). \quad (A4)$$

The reduction from equation (A3) to (A4) uses the property of $C^k(lm,l'm')$ that non-zero values are available only when the following three equations are satisfied:

$$s = m - m', \quad (A5\text{-}1)$$

$$k + l + l' = \text{even}, \quad (A5\text{-}2)$$

and

$$|l - l'| \leq k \leq l + l'. \quad (A5\text{-}3)$$

Whereas the equations required to derive equation (A4), *i.e.*, equations (A3) and (A5), correspond to equations (42) and (10) to (12) in the original XAO paper (Tanaka *et al.*, 2008), that paper did not include an explicit equation, that is, equation (A4). For the implementation of an efficient source code, exclusion of the summation over *s* in equation (A3) is preferable. *REFOWF* (Sakakura, 2017) implements equation (A4). For values of $C^k(lm,l'm')$, see Chapter 5 of Condon & Shortley (1951).

In the case of Ti-$3d^1$ with $l = 2$, equation (A4) becomes:

$$f_{m,m'}(\mathbf{h}) = \sum_{k=0}^{4} i^k \sqrt{4\pi(2k+1)} \langle j_k(Q) \rangle Y_{k,m-m'}^*(\theta',\phi') C^k(2m,2m'). \quad (A6)$$

Since $C^k(2m,2m') = 0$ for odd $k$, $f_{m,m'}$ is only composed of the even terms of $\langle j_k \rangle$.

Since Fourier transformation converts $\rho$, i.e., $\psi^*\psi$, to $f$, the most straightforward proof of vanishing $\langle j_{k\neq 0} \rangle$ for half- or full-filled subshells starts from calculation of $\psi^*\psi$. According to the spherical harmonic addition theorem (see Chapter 3 of Condon & Shortley (1951))

$$P_l(\cos\omega) = \frac{4\pi}{2l+1}\sum_{m=-l}^{l} Y_{l,m}^*(\theta,\phi)Y_{l',m'}(\theta',\phi'), \quad (A7)$$

where, $\omega$ is the angle between two normal vectors expressed in polar coordinates and satisfies $\cos\omega = \cos\theta\cos\theta' + \sin\theta\sin\theta'\cos(\phi - \phi')$, and $P_l(x)$ is the Legendre polynomials and has the recursion relation (see Chapter 12 of Arfken & Weber (2005))

$$(n+1)P_{n+1}(x) = (2n+1)xP_n(x) - nP_{n-1}(x), \quad (A8)$$

with $P_0(x) = 1$ and $P_1(x) = x$.

For the case of $\psi^*(\mathbf{r})\psi(\mathbf{r})$, $\omega = 0$. Therefore, (A7) becomes

$$\sum_{m=-l}^{l}|Y_{l,m}|^2 = \frac{2l+1}{4\pi}P_l(1) = \frac{2l+1}{4\pi} = (2l+1)|Y_{0,0}|^2. \quad (A9)$$

Since the angular part is $Y_{0,0}$ and $l = 0$, only $k = 0$ is available in equation (A4). Thus, $\langle j_{k\neq 0}\rangle$ vanishes.

Another proof by summing $f_{m,m}$ over $m$ requires the values of $C^k(lm,l'm')$ which are given in Chapter 5 of Condon & Shortley (1951). A special case for $d$ subshell is proven in the following. The scattering factor for a half-filled $d$ subshell is given by

$$\sum_{m=-2}^{2} f_{m,m} = \sum_{k=0}^{4} i^k \sqrt{4\pi(2k+1)}\langle j_k\rangle Y_{k,0}^* \sum_{m=-2}^{2} C^k(m,m), \quad (A10)$$

where, $C^k(m,m)$ corresponds to $C^k(2m,2m)$ but without the trivial quantum number $l = 2$ for simplicity. For $k = 4$, $\sum C^4(m,m) = 0$ (since $\sqrt{441}C^4(m,m)$ for respective $m = -2, -1, 0, 1,$ and 2 are 1, −4, 6, −4, and 1), and for $k = 2$, $\sum C^2(m,m) = 0$ (since $\sqrt{49}C^2(m,m)$ for respective $m = -2, -1, 0, 1,$ and 2 are –2, 1, 2, 1, and –2). Thus, $\langle j_{k\neq 0}\rangle$ vanishes for full- or harf-filled $d$ subshells. For $k=0$, $\sum C^0(m,m) = 5$ (since $C^0(m,m)$ for respective $m = -2, -1, 0, 1,$ and 2 are 1, 1, 1, 1, and 1). Thus, only the spherical component $\langle j_0\rangle$ remains.

**Appendix B. Reason for decrease of $F_{\text{orb}}/F$ by inclusion of anomalous dispersion terms**

By expressing $F$ as $F = \mathbf{A} + \mathbf{B}$, where $\mathbf{A} = \sum(f+f')\exp(2\pi i\mathbf{hr})$ and $\mathbf{B} = \sum if''\exp(2\pi i\mathbf{hr})$, a reflection with a very small $F$ roughly satisfies $|\mathbf{A} + \mathbf{B}| = 0$. This is identical to the condition $\mathbf{A} = -\mathbf{B}$, i.e., $\sum(f+f')\exp(2\pi i\mathbf{hr}) = -\sum if''\exp(2\pi i\mathbf{hr})$. It can be seen that $\mathbf{A} = -\mathbf{B}$ can hardly be satisfied, since the only difference between $\mathbf{A}$ and $-\mathbf{B}$ is the replacement of $(f+f')$ with $-if''$. For centric space groups including *Pnma* of YTiO$_3$, satisfaction of $\mathbf{A} = -\mathbf{B}$ is formally forbidden, since atoms locate at $\mathbf{r}$ and $-\mathbf{r}$ simultaneously, and $\mathbf{A}$ is a real number and $\mathbf{B}$ is a complex number (note that $\exp(2\pi i\mathbf{hr}) + \exp(-2\pi i\mathbf{hr}) = 2\cos(2\pi\mathbf{hr})$). Although the term $f'$ in $\mathbf{A}$ simply increases or decreases the effective charge contributing to $F$ from $f$ to $f+f'$ for each atom, $f''$ in $\mathbf{B}$ causes phase delay which is uninterpretable as a shift of the effective charges and prevents reflections from satisfying the condition $|\mathbf{A} + \mathbf{B}| = 0$. Therefore, as the imaginary component of the anomalous dispersion term $f''$ increases, $F_{orb}/F$ generally decreases.

**Appendix C. $\sigma(F)$ for Poisson statistics**

Let $I_{count}$ be the counts collected by a detector, and $F_{count}$ be the square root of $I_{count}$. Propagation of error using the first-order Taylor series leads to the following equation (Wilson, 2006):

$$\sigma(F_{count}) = \sqrt{\left(\frac{\partial I_{count}}{\partial F_{count}}\right)^2 \sigma^2(I_{count})} = \frac{\sigma(I_{count})}{2\,F_{count}}. \qquad (C1)$$

Although the first equal sign in equation (C1) should really be "≈," neglecting the higher-order Taylor series is common in applications of error propagation, and the present paper uses the equal sign for simplicity. If $I_{count}$ purely obeys Poisson statistics, $\sigma(I_{count})$ is given by $F_{count}$. Thus, equation (C1) becomes $\sigma(F_{count}) = 0.5$. By introducing an equation connecting $F_{count}$ and $F_{obs}$ with $F_{count} = k^0_{sc} F_{obs}$, where $k^0_{sc}$ is a scaling factor, $\sigma(F_{obs})$ is obtained:

$$\sigma(F_{obs}) = \frac{\sigma(F_{count})}{k^0_{sc}} = \frac{\sigma(I_{count})}{2\,k^0_{sc}\,F_{count}}. \qquad (C2)$$

Putting the result for Poisson statistics, i.e., $\sigma(F_{count}) = 0.5$, into equation (C2) yields $\sigma(F_{obs}) = 0.5/k^0_{sc}$.

**Appendix D. Definition of parameters and decomposition of $k^0{}_{sc}$**

$$F_{count} = k^0{}_{sc} F_{obs} = k_{sc} (Lp\, A\, Y\, O)^{1/2} F_{obs} \qquad (D1)$$

$$k^0{}_{sc} = k_{sc} (Lp\, A\, Y\, O)^{1/2} \qquad (D2)$$

$Lp$: Lorentz-polarization factor

$A$: Absorption factor

$Y$: Extinction factor

$O$: Non-corrected other factors

For discussion of effective parameters increasing $k^0{}_{sc}$, dependency on crystal radius **r** for spherical crystals is briefly reviewed. Whether $k_{sc}$ is proportional to $V_{crystal}$ or $V_{crystal}{}^{1/2}$ is determined by the coherency of the X-rays and the crystallinity of the specimen, where $V_{crystal}$ is the volume of the crystal. If all the diffracted beams interfere according to the superposition of waves, $k_{sc}$ is proportional to $V_{crystal}$ (note that the superposition of waves is $\boldsymbol{F}_{crystal} = N_{cell}\, \mathcal{F}\, [\rho_{cell}]$, and $N_{cell}$ is given by $V_{crystal}/V_{cell}$. Here, $\boldsymbol{F}_{crystal}$, $N_{cell}$, and $\rho_{cell}$ are, respectively, the total diffraction amplitude from the whole crystal, the number of cells in the crystal, and the EDD in a unit cell), and if the diffracted beams from the spatially separated cells do not interfere with each other, $k_{sc}$ is proportional to $V_{crystal}{}^{1/2}$ (note that the addition of the intensity $|\boldsymbol{F}_{crystal}|^2 = N_{cell}\, |\mathcal{F}\, [\rho_{cell}]|^2$ is satisfied for the present case). For a spherical crystal, $V_{crystal}$ is $4\pi r^3/3$, where $r$ is the crystal radius. Considering the dependency of $A$ on $r^n$ for a spherical crystal summarized in Appendix E, $k^0{}_{sc}$ is proportional to $r^n I_0{}^{1/2} t^{1/2}$ for $n$ from 0.5 to 3. Here, $I_0$ and $t$ are, respectively, the incident intensity and the measurement time. Therefore, enlargement of $r$, $I_0$, and $t$ is effective for increasing $k^0{}_{sc}$.

**Appendix E. Dependency on $r^n$ in absorption factor for spherical crystal**

The absorption factor $A$ is defined by

$$A = \frac{1}{V_{crystal}} \int \exp\{-\mu(t_1 + t_2)\}\, dv, \qquad (E1)$$

where $V_{crystal}$, $\mu$, $t_1$, and $t_2$ are, respectively, the crystal volume, linear attenuation coefficient, path length from the inlet crystal surface to the finite volume $dv$, and path length from the finite volume $dv$ to the outlet crystal surface. Therefore, $A$ physically corresponds to the proportion of effective crystal volume diffracting the beam. The present appendix reviews the dependency of $A_{sph}$ on $r^n$, where $A_{sph}$ is $A$ for a spherical crystal.

In equation (E1), the term $\int \exp\{-\mu(t_1+t_2)\}dv$ is the effective volume, $V_{effective}$, and $V_{crystal}$ is given by $4\pi r^3/3$. Thus, equation (E1) leads to

$$V_{effective} = \frac{4\pi r^3}{3} A_{sph}. \qquad (E2)$$

If $\mu = 0$ (i.e., $A_{sph} = 1$), $V_{effective}$ has a dependency on $r^3$. If $\mu$ is very large, only the surface region of the sphere can diffract the beams, and the beams cannot be transmitted through the crystal. Therefore, $V_{effective}$ is proportional to $r^2$. This situation is similar to moonlight reaching the Earth. When the diffraction angle $\theta$ is 0, this is the case of the new moon, and the diffracted beam from the one-dimensional ring-shaped region can reach the detector. Thus, $V_{effective}$ is proportional to $r^1$.

Since the dependency of $V_{effective}$ on $r^n$ varies from $r^1$ to $r^3$, the dependency of $A_{sph}$ on $r^n$ varies from $r^{-2}$ to $r^0$.

The following is an additional topic on the choices for the series interpolating $A^*_{sph}$ ($\equiv 1/A_{sph}$). Since enlargement of $\mu$ and $r$ has the same effect on $A^*_{sph}$, $A^*_{sph}$ is generally tabulated for discrete $\mu r$, and interpolation by the following equation is proposed (Dwiggins, 1975a, 1975b)

$$A^*_{sph}(\mu r) = \exp\{\sum_{m=1}^{M} K_m (\mu r)^m\}, \qquad (E3)$$

where $K_m$ are the coefficients to $(\mu r)^m$. However, as shown above, the dependency of $A_{sph}$ on $r^n$ varies from $r^{-2}$ to $r^0$, and the dependency of $A^*_{sph}$ on $r^n$ varies from $r^0$ to $r^2$. Therefore, the dependency of $A^*_{sph}$ on $r^n$ is not exponential. Thus, finite series expansion on $r$ is more advantageous than equation (E3) in terms of precision.

**Appendix F. Derivation of equation (3) from equation (2)**

When equation (F1) is introduced, $\sigma(I_{count})$ is given as equation (F2):

$$I_{\text{count}} = I_{\text{corr}} M_{\text{sc}} \tag{F1}$$

$$\sigma(I_{\text{count}}) = \left( M_{\text{sc}}^2 \, \sigma^2(I_{\text{corr}}) + \left(\frac{I_{\text{count}}}{M_{\text{sc}}}\right)^2 \sigma^2(M_{\text{sc}}) \right)^{\frac{1}{2}}. \tag{F2}$$

Using the equations $\sigma(F_{\text{count}}) = \sigma(I_{\text{count}})/(2\,F_{\text{count}})$ and $\sigma(F_{\text{obs}}) = \sigma(F_{\text{count}})/k^0{}_{\text{sc}}$, $\sigma(F_{\text{obs}})$ is derived as:

$$\sigma(F_{\text{obs}}) = \frac{\sigma(I_{\text{count}})}{2\,k_{sc}^0\,F_{\text{count}}} = \left( \left(\frac{M_{\text{sc}}\,\sigma(I_{\text{corr}})}{2\,k_{sc}^0\,F_{\text{count}}}\right)^2 + \left(\frac{\sigma(M_{\text{sc}})}{2\,M_{\text{sc}}}\right)^2 \left(\frac{F_{\text{count}}}{k_{sc}^0}\right)^2 \right)^{\frac{1}{2}}. \tag{F3}$$

By defining $\sigma'(F_{\text{obs}})$ as:

$$\sigma'(F_{\text{obs}}) = \frac{M_{\text{sc}}\,\sigma(I_{\text{corr}})}{2\,k_{sc}^0\,F_{\text{count}}}, \tag{F4}$$

equation (D3) is transformed to:

$$\sigma(F_{\text{obs}}) = \left( \left(\sigma'(F_{\text{obs}})\right)^2 + \left(\frac{\sigma(M_{\text{sc}})}{2\,M_{\text{sc}}}\right)^2 F_{\text{obs}}^2 \right)^{\frac{1}{2}}. \tag{F5}$$

## Appendix G. Systematic bias in refinement on $F^n$

The following precondition is introduced. Observation of $y$ is generated by the generator $f^0(\mathbf{p})$ with a random error for which the standard deviation is $\sigma(y)$, where $\mathbf{p}$ is the parameter vector for $f^0$. (The superscript 0 indicates that $f^0$ has no systematic error.) Then, the minimization function $S(y)$ is given by:

$$S(y) = \sum \left\{ \frac{y - f^0(\mathbf{p})}{\sigma(y)} \right\}^2. \tag{G1}$$

By substituting $F^n$ into $y$, $S(F^n)$ becomes:

$$S(F^n) = \sum \left\{ \frac{F^n - \left(f^0(\mathbf{p})\right)^n}{\sigma(F^n)} \right\}^2. \tag{G2}$$

Since $\sigma(F^n)$ is given as $(\partial(F^n)/\partial F)\sigma(F)$ by error propagation, equation (G2) is transformed to

$$S(F^n) = \sum \left\{ \frac{F^n - \left(f^0(\mathbf{p})\right)^n}{\frac{\partial(F^n)}{\partial F}\sigma(F)} \right\}^2. \tag{G3}$$

Let us regard $F$ as the $x$-value and $F^n$ as the $y$-value. Then, the two points $(F, F^n)$ and $(f^0(\mathbf{p}), (f^0(\mathbf{p}))^n)$ can define $\Delta y$ and $\Delta x$ as $F^n - (f^0(\mathbf{p}))^n$ and $F - f^0(\mathbf{p})$, respectively. The slope $dy/dx$ at point $(F, F^n)$ is given by $\partial(F^n)/\partial F$. Thus, the term $(F^n - (f^0(\mathbf{p}))^n)/(\partial(F^n)/\partial F)$ in equation (G3) corresponds to $\Delta y /(dx/dy)$ and gives an approximation of $\Delta x$. The reason why the approximation of $\Delta x$, namely the approximation of $\Delta F$ in this case, is required for the optimization in $F^n$-space has to do with the fact that the superposition principle of waves, which connects direct and reciprocal space, is defined in $F^1$-space. To solve for $\mathbf{p}$, the conversion of quantities from $F^n$-space to $F^1$-space using $(\partial F^n/\partial p) = (\partial F^n/\partial F)(\partial F/\partial p)$ is required for a secular equation with any dimension of $n$. Thus, the approximation of $\Delta F$ by $(F^n - (f^0(\mathbf{p}))^n)/(\partial(F^n)/\partial F)$ cannot be avoided.

When $n$ is set to 1, equation (G3) becomes $S(F) = \Sigma\{(F - f^0(\mathbf{p}))/\sigma(F)\}^2$, and $\Delta F$ is given by $F - f^0(\mathbf{p})$ with no approximation error. As $n$ is far from 1, the approximation error grows large. The graphical representation of this is shown in Fig. 4(a).

As Fig. 4(a) shows, the magnitude of the approximation error is also enlarged when $\Delta F$ is large, and it is also confirmable that the magnitude of the approximated $\Delta F_{\text{approximated}}$ ($\equiv (F^n - (f^0(\mathbf{p}))^n)/(\partial(F^n)/\partial F)$) changes when the position of the points $(F, F^n)$ and $(f^0(\mathbf{p}), (f^0(\mathbf{p}))^n)$ are swapped even if the magnitude of $|\Delta F|$ is maintained, since the slope at $(F, F^n)$ changes. This causes a fluctuation of $\mathbf{p}$ in every refinement cycle and makes the refinement unstable.

**Appendix H. The Poisson distribution in $F^1$- and $F^2$-space**

The Poisson distribution $P(y)$ defined for the true expected value $y_0$ is given by equation (H1), and the normal distribution $N(y)$ defined for the true expected value $y_0$ with the standard deviation $\sigma(y)$ is given by equation (H2).

$$P(y) = \frac{y_0^y}{y!} \exp(-y_0) \tag{H1}$$

$$N(y) = \frac{1}{\sigma(y)\sqrt{2\pi}} \exp\left\{-\frac{1}{2}\left(\frac{y-y_0}{\sigma(y)}\right)^2\right\} \tag{H2}$$

The parameter conversion from $y$ to $x$ for a probability density function $Q(y)$ is accomplished by $Q(y)(dy/dx)$. Therefore, the Poisson distribution $P(y)$ in $F^1$-space is defined by $P(y)(2x)$, where $x = F$, and $y = F^2$. Figures 4(b), S7(a), and S7(c) are calculated by $P(F^2)(2F)$. The multiplication factor $2F$ contributes to reduction of the asymmetry of the Poisson distribution $P(y)$ at small $F$, and the normal distribution can provide a good approximation even at small $F$ in $F^1$-space.

**Appendix I. $F_{calc}$-dependent weight**

The weight $w = (1/\delta F_{estimated})^2 = 1/\{a + b(1/3 F^2_{obs} + 2/3 F^2_{calc})\}$ (Wilson, 1976) is one of the most common weights available in systems such as *SHELX* (Sheldrick, 2008). This weight is intended to cancel out the statistical fluctuation in $F^2_{obs}$ by introducing a non-zero $\delta^2 F_{calc}$ and choosing a weight that will satisfy $(\partial w/\partial F^2_{calc}) = 2(\partial w/\partial F^2_{obs})$ (Wilson, 1976). However, the introduction of $\delta F_{calc}$ to explain the statistical fluctuation in $F_{obs}$, i.e., $\delta F_{obs}$, with $\delta F_{calc}$ is not allowed in the least-squares method. As indicated by the fact that the least-squares method is a special case of the maximum likelihood method whose statistical distribution function is a normal distribution (Prince & Collins, 2006), statistical fluctuation is already considered and statistical bias should not be introduced in $F_{calc}$. (If the error distribution is distorted from the normal distribution, the least-squares method should be replaced with the maximum likelihood method.) The introducible source of $\delta F_{calc}$ is a model for the correction of systematic errors in $\delta F_{obs}$ (as represented, for example, by the extinction factor), and this is always done by improving the refinement model by a person refining the data.

By connecting $F_{obs}$ and $F_{calc}$ with $(\partial w/\partial F^2_{calc}) = 2(\partial w/\partial F^2_{obs})$, the weight $(1/\delta F_{estimated})^2 = 1/\{a + b(1/3 F^2_{obs} + 2/3 F^2_{calc})\}$ indicates that twice as large a baseline statistical uncertainty lies in $F^2_{calc}$ than in $F^2_{obs}$ (see Section 2.2.2 for the physical meaning of the term proportional to $F^2$). This is physically inaccessible.

In practice, introduction of the recursive dependency on $F_{calc}$ through the weight prevents maximum use of the experimental data $F_{obs}$, since $F_{calc}$, which is to be optimized from $F_{obs}$, depends on $F_{calc}$ itself.

**Appendix J. Brief review of maximum entropy method**

In Section 2, the equation $\rho(\mathbf{r})$ was developed as:

$$\rho(\mathbf{r}) = (1/V_{cell})\sum_{\mathbf{h}}\{\mathbf{F}(\mathbf{h})\exp(-2\pi i \mathbf{hr})\}$$

$$= (1/V_{cell})\sum_{\mathbf{h}}\{\sum_{atom} f_{atom}(\mathbf{h})\exp(2\pi i \mathbf{hr}_{atom})\}\exp(-2\pi i \mathbf{hr})$$

$$= (1/V_{cell})\sum_{atom}\sum_{\mathbf{h}} f_{atom}(\mathbf{h})\exp\{-2\pi i \mathbf{h}(\mathbf{r}-\mathbf{r}_{atom})\}$$

$$= (1/V_{cell})\sum_{atom} \rho_{atom}(\mathbf{r}) \qquad (J1)$$

, where $\mathbf{F}(\mathbf{h}) = \sum_{atom} f_{atom}(\mathbf{h})\exp(2\pi i \mathbf{hr}_{atom})$ and $\rho_{atom}(\mathbf{r}) = \sum_{\mathbf{h}} f_{atom}(\mathbf{h})\exp\{-2\pi i \mathbf{h}(\mathbf{r}-\mathbf{r}_{atom})\}$. Equation (J1) proves that when a certain range of $\mathbf{F}(\mathbf{h})$, *i.e.*, information of $f_{atom}(\mathbf{h})$, is truncated, no waves which have maxima from the nucleus at a certain $d$ is not superimposed to the $\rho_{atom}$. MEM is the method finding out the minima of $\chi^2$ enforcing the maximization of information entropy $S$ by introducing Lagrangian multiplier $\lambda$ as:

$$L(\lambda) = \chi^2 - 1/\lambda\ S. \qquad (J2)$$

Here, $\chi^2 = \Sigma((F_{obs} - F_{calc})/\sigma(F_{obs}))^2$, and $S = -\sum_j n_j \ln\left(\frac{n_j}{n'_j}\right)$, where, $n_j$ is the number of electrons in the *j*-th voxel in a unit cell to be refined in the present iteration, and $n'_j$ is the number of electrons in the *j*-th voxel refined in the previous iteration. In most literatures, equation (J2) is defined in a different form such as (*e.g.*, eq. (5.52) in the Coppens's book (1997)):

$$L'(\lambda) = S - \lambda\ \chi^2. \qquad (J3)$$

However, for comparison to other methods such as maximum likelihood and least-squares, which searches the minima of the method-specific minimization functions, inverting the sign of the equation (J3) as (J2) and swapping the problem from finding maxima of entropy (eq. J3) to finding minima of $\chi^2$ (eq. J2) is preferable. Note that the maximization of entropy simply keeps $\rho$ to the $\rho$ for the previous iteration and cannot refine $\rho$ at all. This behavior of the entropy term $S$ is explained by the following. According to the third principle of thermodynamics, the entropy $S$, which is also known as Kullback–Leibler divergence in mathematical statistics (Kullback and Leibler, 1951), $S \geq 0$ is satisfied (*e.g.*, Section 44-6 of Feynman *et al.* (1965)). Since $S = 0$ is realized when $n_j/n'_j = 1$ is satisfied for all

$j$, the term $S$ act as constraint so as not to change the distribution from the previous iteration of $n'_j$. Putting (J2) into $\partial L/\partial n_j = 0$, which is satisfied for minima, leads to $\lambda(\partial \chi^2/\partial n_j) = (\partial S/\partial n_j)$. By taking exponent of this equation, $n_j$ is derived as:

$$n_j = (n'_j/Z)\exp\{-\lambda(\partial \chi^2/\partial n_j)\}, \qquad (J4)$$

where, $Z$ is the normalization factor to meet $\sum n_j = F(0,0,0)$. The physical meaning of equation (J4) is that the number of electrons in a voxel at the last iteration $n'_j$ is scaled by the weight $(1/Z)\exp\{-\lambda(\partial \chi^2/\partial n_j)\}$. For searching minimum $\chi^2$, $n_j$ should be increased along the direction negating $(\partial \chi^2/\partial n_j)$ from the local point of view (see any introductions for steepest descent method). Therefore, MEM amplify the magnitude of parameter shift calculated from $-\lambda(\partial \chi^2/\partial n_j)$ with exponential dependency. Among the voxels with the same magnitude of $\exp\{-\lambda(\partial \chi^2/\partial n_j)\}$, voxels with larger $n'_j$ in the previous iteration have larger $n_j$, and for voxels with the same magnitude of $n'_j$, voxels with larger magnitude of $\exp\{-\lambda(\partial \chi^2/\partial n_j)\}$ have larger $n_j$.

Therefore, for clear detection of anisotropies by MEM, realization of larger $n_j$ and larger $-\lambda(\partial \chi^2/\partial n_j)$ in the refinement is the significant condition. In general, when reflections in a certain region of $d$ is truncated, it is expectable that large $-\lambda(\partial \chi^2/\partial n_j)$ cannot be obtained to describe the anisotropy at the truncated $d$ from the nucleus.

Another important point in MEM is that $\chi^2$ is generally provided by vector $\boldsymbol{F}$ instead of scalar $F$ (see *e.g.*, eq. (5.50) in the Coppens's book (1997)):

$$\chi^2 = \sum(\boldsymbol{F}_{obs} - \boldsymbol{F}_{calc})^2/\sigma^2(\boldsymbol{F}_{obs}) . \qquad (J5)$$

Therefore, the resulting EDDs strongly dependent on the model used to attach the complex phase to scalar $F$.

# Supporting Information

## S1. Supplemental files

The following files are supplemental attachments: HKL files which are used as the input of crystal structure analysis in the present study, Fobs-Fcalc files which are the output of the refined structure factors, and CIF files. HKL and Fobs-Fcalc files are headed by one comment line starting with the character '#'. The comment lines explain the data types aligned in the columns in each data file. Since all the supplemental files are written in ASCII characters, superscripts and subscripts are not available. In the following, italic strings are the notation used in the ASCII files.

The data types in HKL files are '*# h k l Fobs SigFobs Tbar*'. Here, $Fobs = (Lp\,A)^{-1} F'_{count}$, where $F'_{count}$ is $F_{count}$ after subtraction of background counts and is scaled using the monitored counts for the incident. *SigFobs* is also scaled $\sigma(F_{obs})$ using the monitored counts. *Tbar* corresponds to '*the absorption-weighted mean path length through the real crystal* (Becker & Coppnes, 1974*a*)' in cm.

Filenames and their corresponding datasets are as follows.

- Fobs_Orbital_3bs.hkl: hkl file for refinement by OWM.
- Fobs_MDA.hkl: hkl file for MDA.
- Fobs_Non-MDA.hkl: hkl file for non-MDA.
- Fobs_APD.hkl: hkl file for APD.
- Fobs_SD.hkl: hkl file for SD.

## S1.1.1. Fobs-Fcalc files

The data types are '*# h k l F0obs Re(F0calc) Im(F0calc) Re(Forb) Im(Forb) sinTh/L Yext sqrt(wgt) SigFobs Re(Fcalc) Im(Fcalc) ksc*,' where *Yext* is the *Y* defined in the present paper (see Appendix D), and *sqrt(wgt)* is calculated by sqrt(1/(*SigFobs*^2+0.000009*Fobs*^2)). *ksc* is calculated by, *e.g.*, 26.619/sqrt(*Yext*) in the refinement by OWM. The functions *Re* and *Im* extract the real and imaginary parts of complex numbers, respectively.

The following is the list of Fobs-Fcalc files:

- Orbital_3bs.out: structure factors after refinement by OWM.
- MDA.out: for MDA.
- Non-MDA.out: for non-MDA.
- APD.out: for APD.

- SD.out: for SD.

### S1.1.2. CIF files

The following are the notes for the attached CIF file. (A) Since the present version of *REFOWF* has no source code that calculates bond lengths and angles, the sections corresponding to bond lengths and angles are filled with '?'. (For bond lengths and angles, one can calculate bond lengths and angles by recent crystal structure visualization programs such as *VESTA* (Momma & Izumi, 2011) by giving a CIF file and clicking on the bonds and angles one wants to know.) (B) As included in the body of the present paper, structure factors were not averaged among the symmetry equivalents. Therefore, the terms *'_diffrn_reflns_number,' '_reflns_number_total,' '_reflns_number_gt,'* and *'_refine_ls_number_reflns,'* which correspond to the number of reflections used in the refinement, are not given by the number of independent reflections.

The following is the list of CIF files:

- Orbital_3bs.cif
- MDA.cif
- Non-MDA.cif
- APD.cif
- SD.cif

### S2. Supplemental figures

The figures referred to as Fig. S* in the text are the following, where * is a wildcard.

**Figure S1** 2D representations of residual EDDs shown in Fig. 12. The contour interval is 0.2 eÅ$^{-3}$. The levels of positive, zero and negative contours are shown by red, dark orange, and green lines, respectively. The described 2D planes are defined by the four density maxima of the lobes of the Ti-$3d^1$ orbital in (a) and (b), the *xy*-plane passing through Y in (c) and (d), and the plane defined by Ti, O1$_x$ and O2$_z$ in (e) and (f).

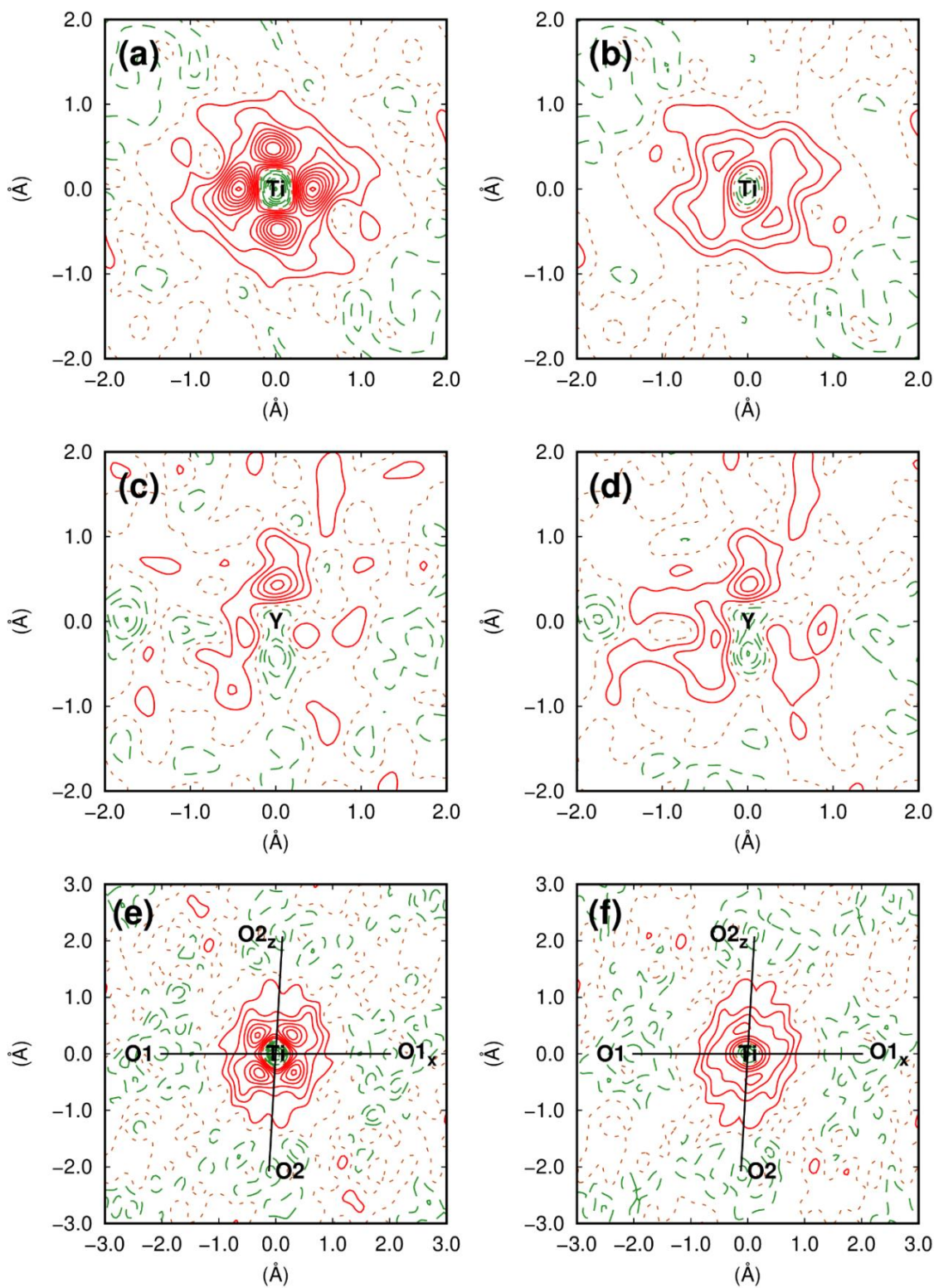

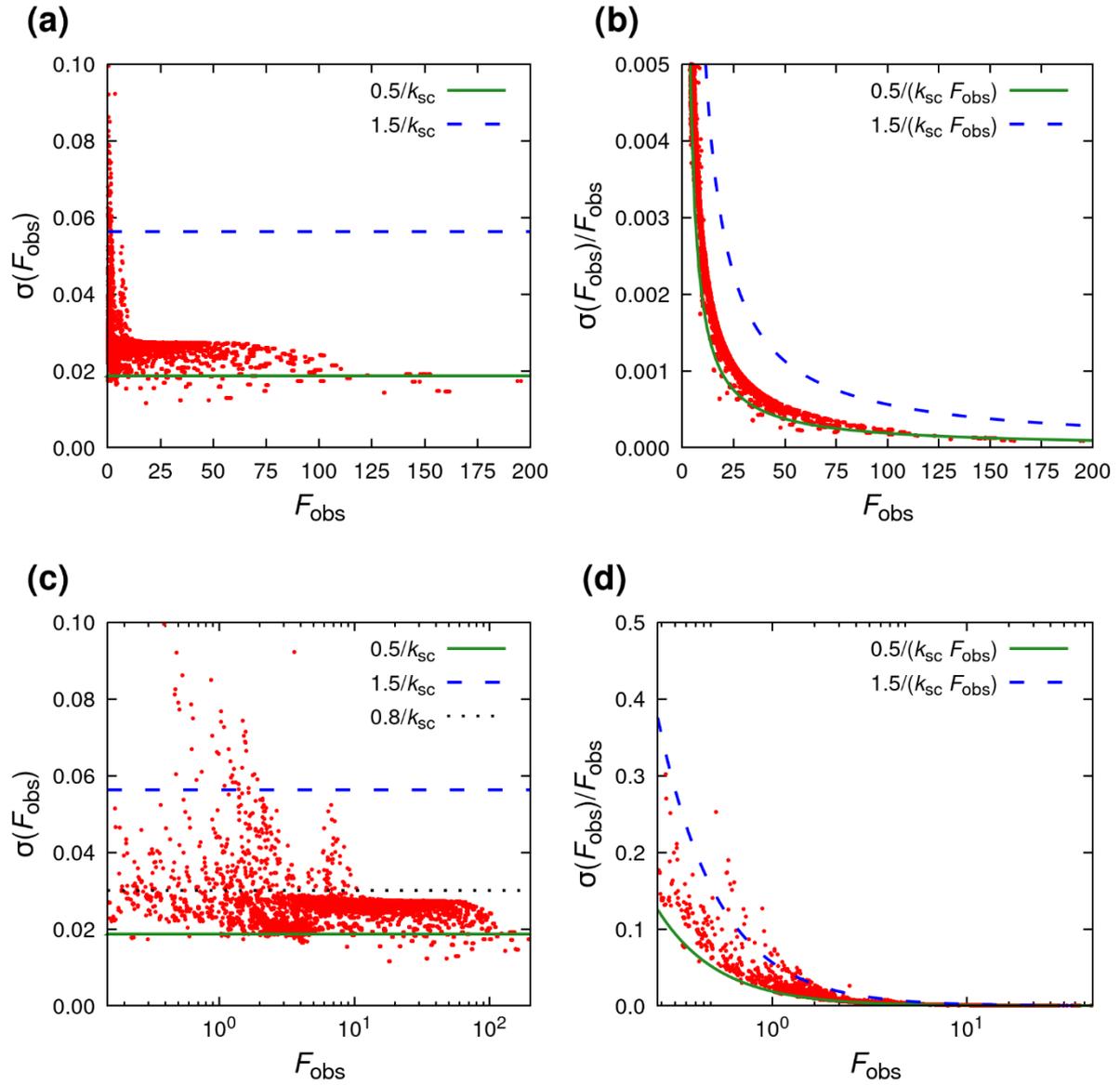

**Figure S2** $\sigma(F_{obs})$ and $\sigma(F_{obs})/F_{obs}$ versus $F_{obs}$. The left panels (a) and (c) are for $\sigma(F_{obs})$, and the right panels (b) and (d) are for $\sigma(F_{obs})/F_{obs}$. In the bottom panels (c) and (d), the horizontal scale is logarithmic.

**Figure S3** 2D representations of residual EDDs shown in Figs. 7(b) and 7(c). The contour interval and the definition of the described 2D planes are the same as Fig. S1. The left panels correspond to Fig. 7(b) for MDA, and the right panels correspond to Fig. 7(c) for non-MDA.

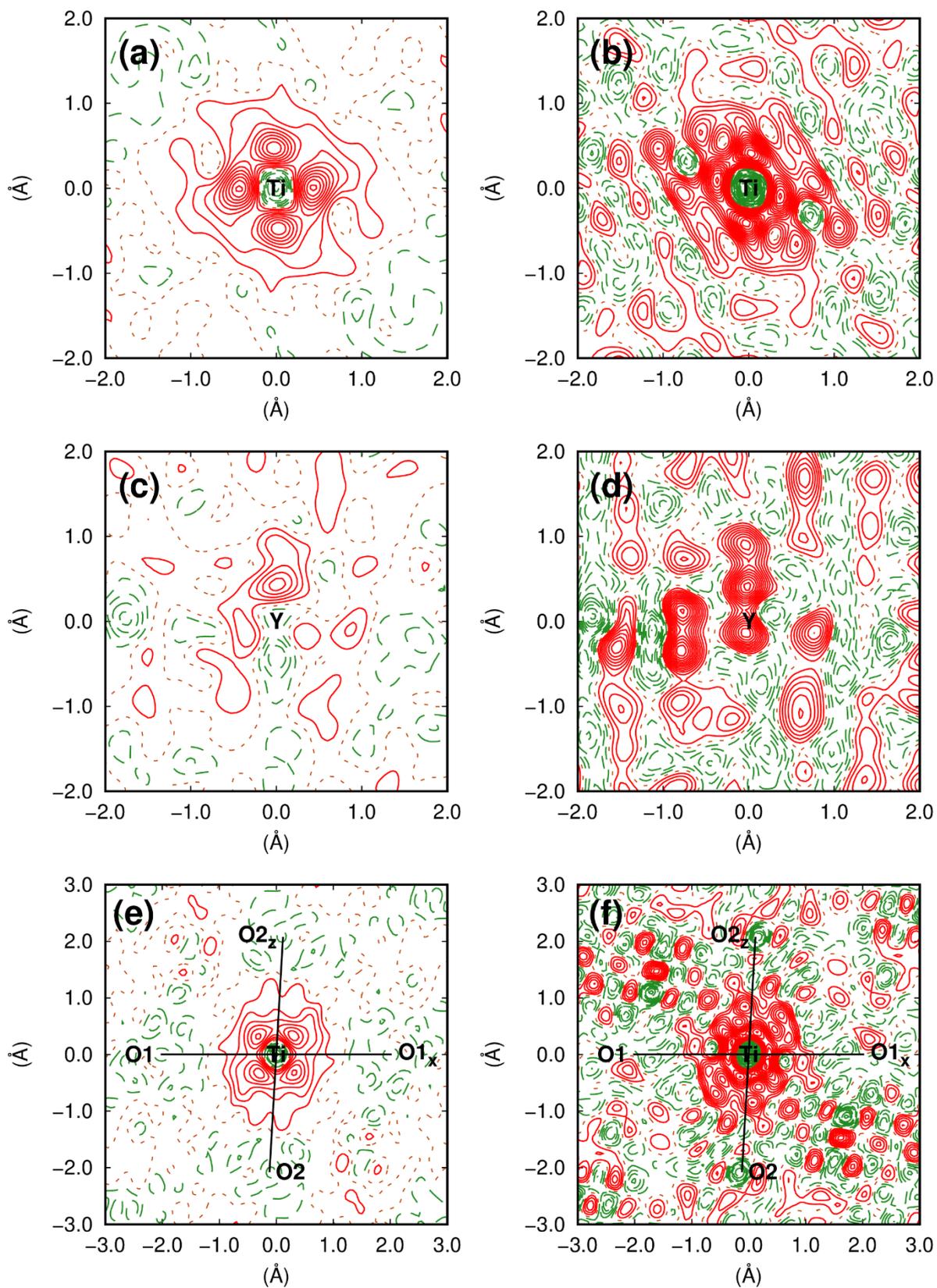

**Figure S4** 2D representations of residual EDDs shown in Figs. 9(b) and 9(c). The contour interval and the definition of the described 2D planes are the same as Fig. S1. The right panels correspond to Fig. 9(b) for the APD detector, and the left panels correspond to Fig. 9(c) for an SD.

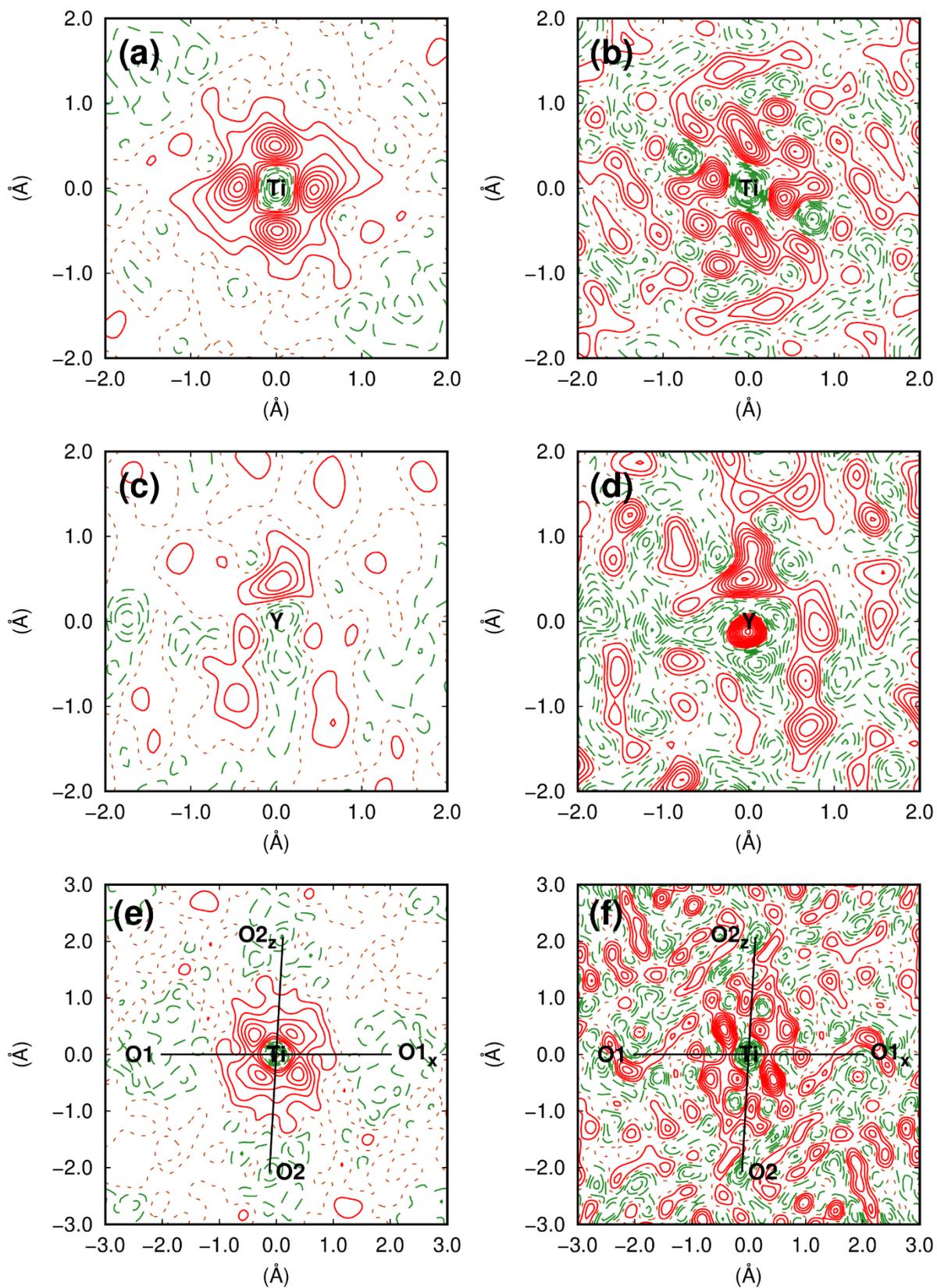

**Figure S5** EDDs obtained by Fourier synthesis with $\sin\theta/\lambda$-cut and $F^0_{min}$-cut for the same dataset as Fig. 2. The ranges of $\sin\theta/\lambda$ are different in columns, and the magnitudes of $F^0_{min}$ are different in rows. From the left, the ranges of $\sin\theta/\lambda$ are $0.0 < \sin\theta/\lambda < 2.0$ Å$^{-1}$, $0.0 < \sin\theta/\lambda < 1.2$ Å$^{-1}$, and $1.2 < \sin\theta/\lambda < 2.0$ Å$^{-1}$. From the top, the magnitudes of $F^0_{min}$ are 0, 5, 13, and 20. $N_{refl}$ shown in each figure indicates the number of reflections used for the Fourier synthesis including $F^0(0,0,0)$.

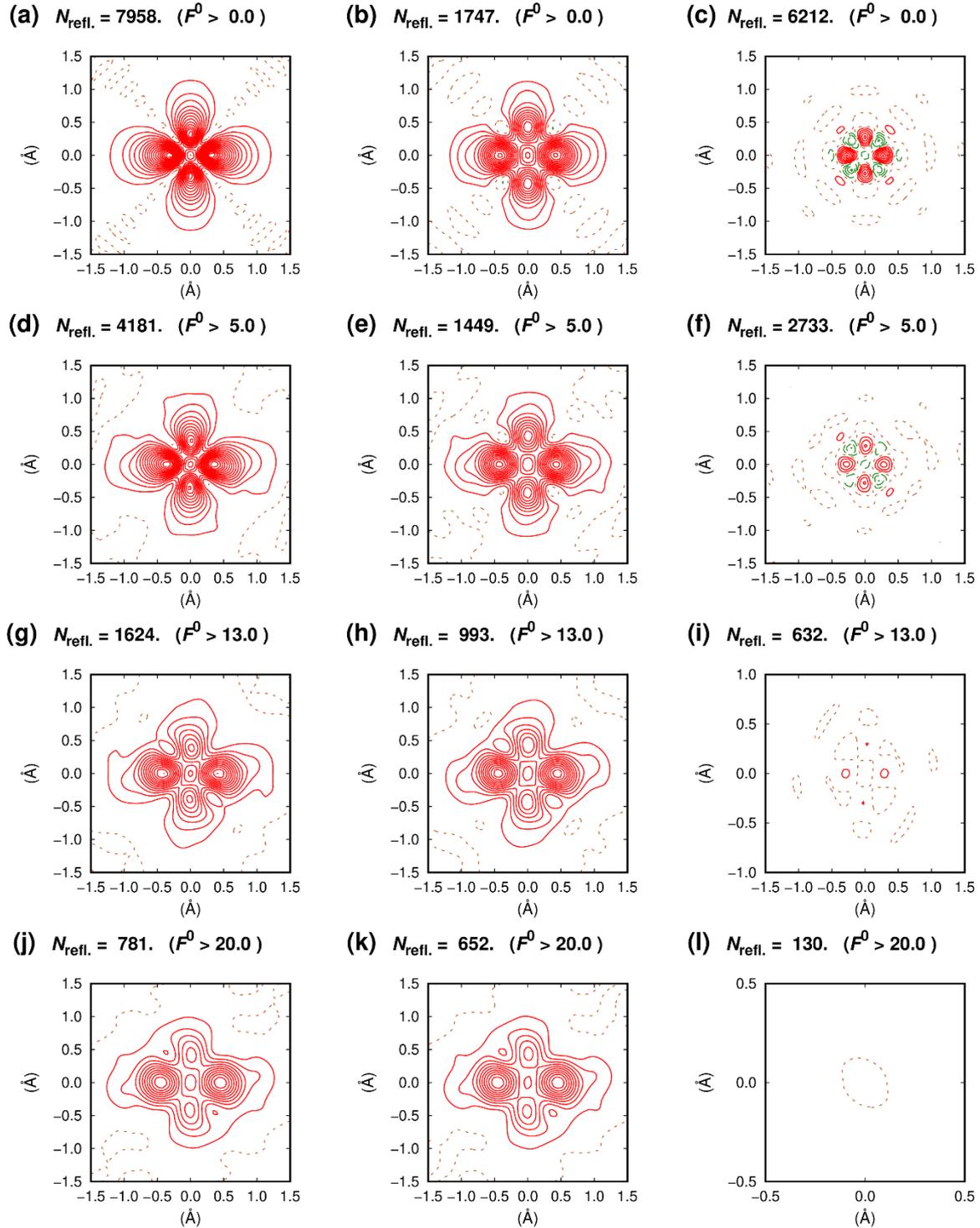

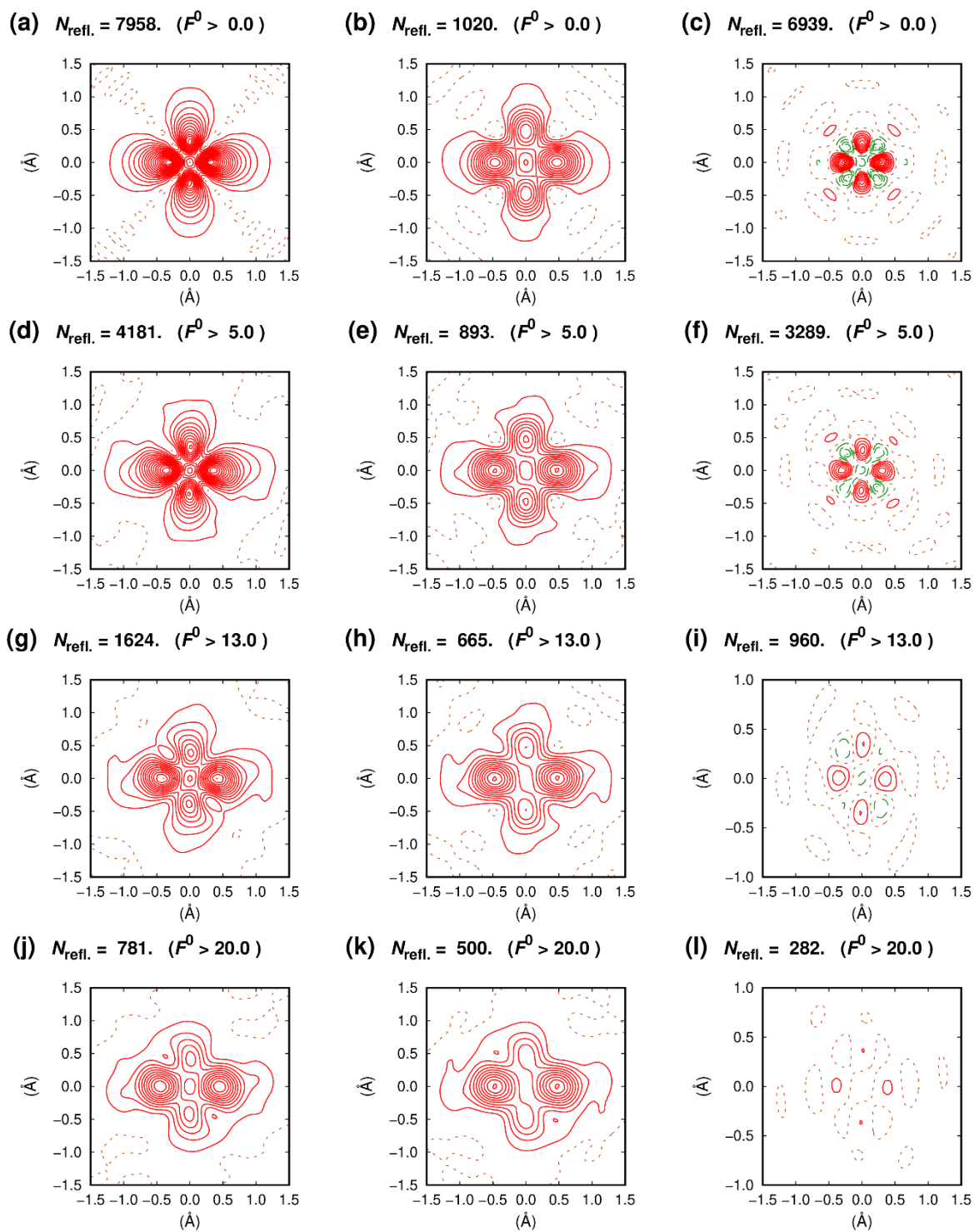

**Figure S6** EDDs obtained in the same manner as Fig. S5 with different ranges of sin$\theta$/$\lambda$-cut to Fig. S5. The threshold sin$\theta$/$\lambda$ dividing the lower- and higher-angle reflections is set to 1.0 Å$^{-1}$. From the left, the ranges of sin$\theta$/$\lambda$ are: 0.0 < sin$\theta$/$\lambda$ < 2.0 Å$^{-1}$, 0.0 < sin$\theta$/$\lambda$ < 1.0 Å$^{-1}$, and 1.0 < sin$\theta$/$\lambda$ < 2.0 Å$^{-1}$.

**Figure S7** The Poisson distribution in $F^1$- and $F^2$-space. (a) and (c) are for $F^1$-space, and (b) and (d) are for $F^2$-space. (a) and (b) are the same events for $I_0 = 2$ counts, and (c) and (d) are the same events for $I_0 = 25$ counts. In $F^1$-space, the Poisson distribution is still well-approximated by a normal distribution with $\sigma(F_{count}) = 0.5$ even for the case of 2 counts, therefore $F_{count} = \sqrt{2}$. For the larger count case of 25 counts, the approximation becomes more precise.

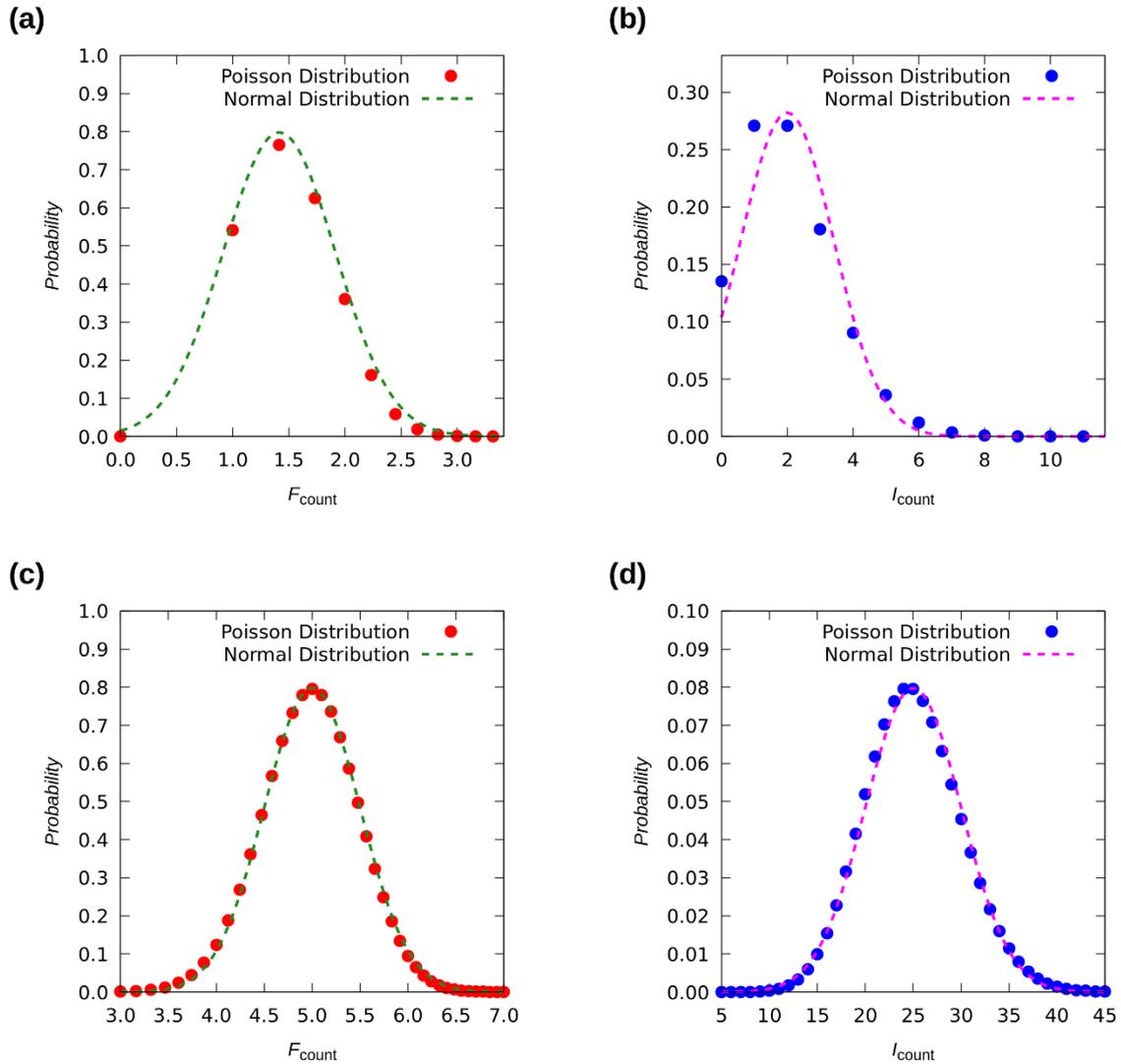

**Figure S8** Spherical single crystal of YTiO$_3$.

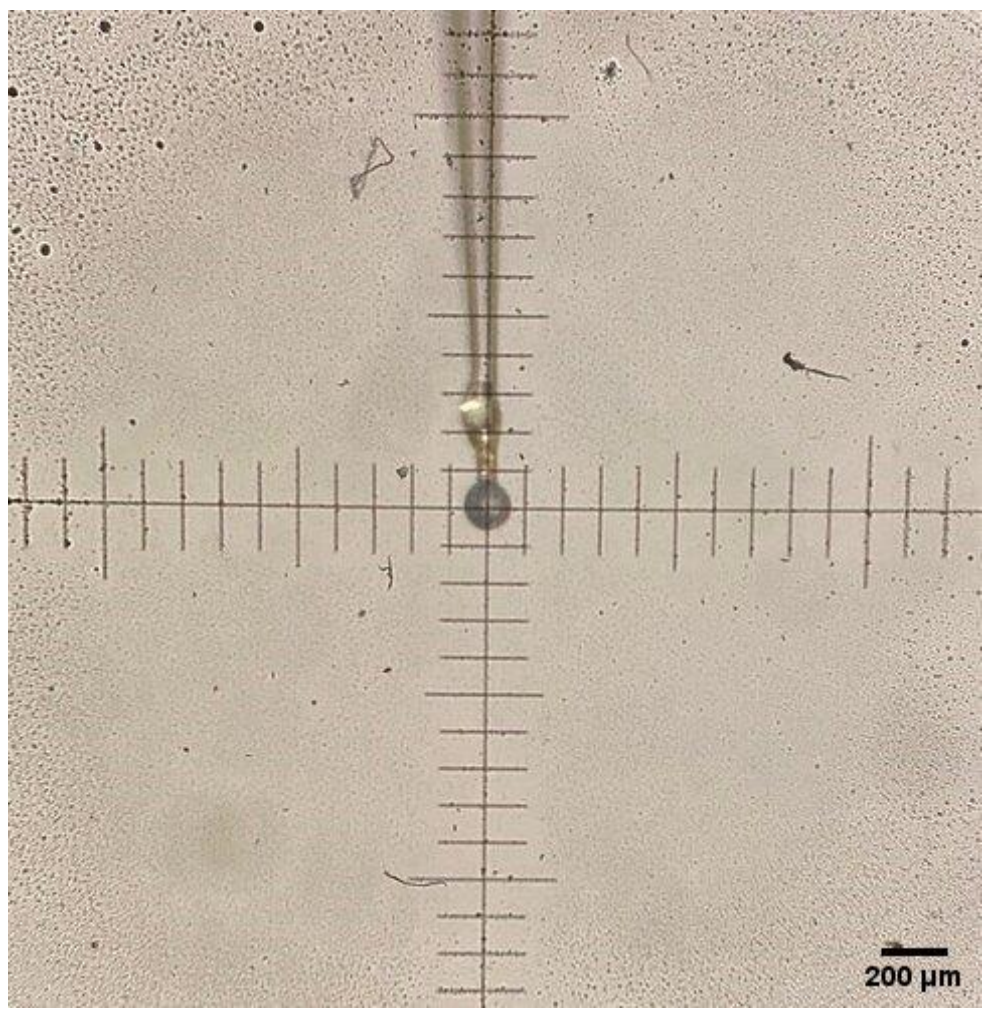